\documentclass[3p,11pt]{elsarticle}
\usepackage{lineno,hyperref}
\modulolinenumbers[1]
\usepackage{geometry}
\usepackage{graphicx}
\usepackage[usenames,dvipsnames]{color}
\usepackage{color}
\usepackage[english]{babel}
\usepackage{epstopdf}
\usepackage{mathtools}
\usepackage{latexsym}
\usepackage{hyperref}
\usepackage{url}
\usepackage{amstext}
\usepackage{amssymb}
\usepackage{amsmath}
\usepackage{pifont}
\usepackage{siunitx} 
\usepackage{url}
\hypersetup{
    colorlinks=true,
    linkcolor=blue,
    filecolor=magenta,
    urlcolor=blue,
}
\usepackage{times}
\usepackage{hyperref}
\usepackage{url}
\usepackage{amstext}
\usepackage{amssymb}
\usepackage{wrapfig}
\usepackage{float}
\usepackage{tikz}
\usepackage{pgfplots}
\usepackage{caption}
\usepackage{subcaption}
\usepackage[section]{placeins}

\usepackage{lineno}

\linenumbers

\begin{document}

\begin{frontmatter}

    \title{Bayesian Approaches for
Revealing Complex Neural Network Dynamics in Parkinson’s Disease}

\author[inst1]{Hina Shaheen\corref{cor1}}
\ead{hina.shaheen@umanitoba.ca}
\author[inst1]{Roderick Melnik}
\ead{rmelnik@wlu.ca}

\cortext[cor1]{Hina Shaheen}
\address[inst1]{Faculty of Science, University of Manitoba, Winnipeg, MB R3T 2N2, Canada}
\address[inst2]{MS2Discovery Interdisciplinary Research Institute, Wilfrid Laurier University, Waterloo, ON N2L 3C5, Canada}

\begin{abstract}
Parkinson’s disease (PD) belongs to the class of
neurodegenerative disorders that affect the central nervous system. It is
usually defined as the gradual loss of dopaminergic neurons in the substantia nigra pars compacta, which causes both motor and non-motor symptoms. Understanding the neuronal processes that underlie PD is critical for creating successful therapies. This study combines machine learning (ML), stochastic modelling, and Bayesian inference with connectomic data to analyze the brain networks involved in PD. We use modern computational methods to study large-scale neural networks to identify neuronal activity patterns related to PD development. We aim to define the subtle structural and functional connection changes in PD brains by combining connectomic with stochastic noises. Stochastic modelling approaches reflect brain dynamics' intrinsic variability and unpredictability, shedding light on the origin and spread of pathogenic events in PD. We employ a novel hybrid model to assess how stochastic noise impacts the cortex-basal ganglia-thalamus (CBGTH) network, using data from the Human Connectome Project (HCP). Bayesian inference allows us to quantify uncertainty in model parameters, improving the accuracy of our predictions. Our findings reveal that stochastic disturbances increase thalamus activity, even under deep brain stimulation (DBS). Bayesian analysis suggests that reducing these disturbances could enhance healthy brain states, providing insights for potential therapeutic interventions. This approach offers a deeper understanding of PD dynamics and paves the way for personalized treatment strategies. This is an extended version of our work presented at the ICCS-2024 conference \cite{shaheen2024neural}.

\end{abstract}


\end{frontmatter}

\section{Introduction}
Parkinson's disease (PD) stands as one of the most prevalent neurodegenerative disorders (NDDs), characterized by the progressive loss of dopaminergic neurons in the substantia nigra pars compacta, leading to debilitating motor symptoms such as tremors, rigidity, and bradykinesia \cite{salaramoli2024therapeutic,shaheen2022multiscale}. Despite significant advancements in therapeutic approaches, including pharmacological interventions and DBS, our understanding of the complex interplay between neuronal dynamics, disease progression, and treatment outcomes remains incomplete \cite{shaheen2023bayesian,johnson2024deep}.

Recent years have witnessed a paradigm shift in neuroscientific research, driven by the convergence of computational methodologies, artificial intelligence (AI) techniques, including ML tools, and advancements in neural engineering \cite{peralta2021machine}. Among these approaches, ML holds promise in deciphering intricate patterns within vast datasets, offering insights into disease mechanisms and personalized treatment strategies. Concurrently, DBS has emerged as a potent therapeutic modality, modulating aberrant neuronal circuits to alleviate motor symptoms in PD patients \cite{tai2019machine}. The discipline of ML, which is a subdivision of AI, has experienced rapid growth and has recently impacted medical fields like neurosurgery \cite{peralta2021machine, ghebrehiwet2024revolutionizing,storm2024integrative}. A literature review focusing on the application of ML in DBS has not yet been published despite the field's growing interest in the area.

In parallel, stochastic modelling has gained traction to capture the inherent randomness and complexity of neuronal activity \cite{thieu2022coupled}, shedding light on the dynamic nature of neurological disorders such as PD \cite{oliveira2023machine}. By integrating these diverse methodologies, researchers aim to unravel the underlying mechanisms governing neuronal dysfunction in PD, thereby paving the way for more effective interventions and improved patient outcomes. Furthermore, recent studies employing multiscale mathematical modelling
have highlighted the efficacy of nonlinear reaction-diffusion equations in discerning neuropathological conditions \cite{meier2022virtual}. Notably, connectomic data has revealed the extensive impact of DBS across various cortical and subcortical regions \cite{meier2022virtual}. Discrete brain network models operating in a spatio-temporal domain elucidate the dynamics of model parameters, thereby simulating large-scale brain activity \cite{shaheen2023bayesian,meier2022virtual,seguin2023brain}.

In essence, neurons constitute the fundamental units of our nervous system, with the basal ganglia (BG) comprising three critical nuclei: the subthalamic nucleus (STN), the globus pallidus internus (GPi), and the globus pallidus externus (GPe) \cite{shaheen2022multiscale}. Neurons utilize neurotransmitters for intercellular communication and employ action potentials to transmit signals within the cell upon receiving external stimuli ($I_{app}$). Notably, using a reduced number of neurons, such as $10$ neurons per nucleus, yields similar outcomes to those obtained with $100$ neurons. Thus, each nucleus in our study comprises $10$ cells \cite{shaheen2022multiscale}.  
Importantly, Approximate Bayesian Computation (ABC) is particularly important in brain modelling and stochastic modelling because it allows us to make inferences about complex systems where the likelihood function is unknown or difficult to calculate \cite{shaheen2023bayesian}. Moreover, Bayesian inference in machine learning provides a probabilistic approach to modelling uncertainty by combining prior knowledge with observed data \cite{tyralis2024review}. In the context of brain modelling, such as the CBGTH system, the brain's dynamics are highly nonlinear and influenced by various sources of noise and uncertainty, making traditional likelihood-based methods infeasible. Stochastic models of brain processes, which incorporate random elements like fluctuations in membrane potentials or synaptic transmission, further complicate the ability to derive explicit likelihoods \cite{deco2009stochastic}. In this study, we utilize ABC to explore the posterior distributions of the standard deviations of the membrane potentials ($\sigma$) for key brain regions: the STN, GPe and GPi, and the TH. By utilizing ABC, we bridge the gap between theoretical models and real-world data, providing insights into how brain networks modulate membrane potential variability across different neural regions, which is critical for understanding the underlying mechanisms of NDDs.

In the present study, we adopted a novel co-simulation approach utilizing a modified Rubin-Terman model for subcortical brain regions surrounding the basal ganglia across the entire cerebral hemisphere from our previous study \cite{shaheen2022multiscale}. This approach incorporates stochastic noise, explicitly incorporating a Wiener process, to capture additional variability and complexity in brain dynamics \cite{novelli2024spectral,vashistha2024parapet}. Therefore, we integrate a discrete brain network model for each cortical region, incorporating stochastic noise at the macroscopic scale to better align with experimental data on neuron firing characteristics. Following the strategy outlined in \cite{shaheen2022multiscale}, we explore critical aspects of the model dynamics, including the influence of stochastic noise on healthy and diseased states. Our findings demonstrate that the eigendecomposition of the Laplace operator, incorporating stochastic noise, can predict the collective dynamics of human brain activity at the macroscopic scale \cite{shaheen2022deep}. These findings suggest that the disruption of multivariate connection-wise functional connectivity patterns holds promise for discriminating PD patients based on cognitive status, supporting previous observations of altered functional connectivity associated with cognitive impairment in PD. Our research uncovers significant findings regarding the influence of stochastic noise on brain dynamics. Specifically, we observed that in the presence of stochastic noise, the activity of the thalamus reaches a critical threshold, contrasting with scenarios lacking noise.

Furthermore, our analysis revealed that stochastic noise amplifies the membrane potential of the thalamus, potentially exacerbating brain disease states. This effect of stochastic noise is pronounced, leading to burst oscillations in the membrane potential across all selected regions, even in the presence of DBS. Our study highlights the brain's resilience as it endeavours to maintain a healthy state for a prolonged period following DBS despite stochastic noise. In addition, we found that the posterior distributions of the standard deviations of the membrane potentials ($\sigma$) for the STN, GPe, GPi, and TH, derived using ABC, revealed significant variations across these regions. These variations suggest that different parts of the cortico-basal-thalamic system exhibit distinct patterns of neural variability, which may play a critical role in disease dynamics. Specifically, the posterior distributions indicate that regions such as the STN and GPi, which are central to motor control, show higher fluctuations in membrane potential variability in Parkinsonian states, potentially contributing to the motor symptoms observed in the disease. This insight, derived through the use of ABC, highlights the value of stochastic modelling in understanding how noise and variability in brain networks correlate with neurodegenerative progression, offering new perspectives for diagnostic and therapeutic approaches. We plotted histograms of the posterior distributions to visually assess the model fit. The results showed that our data aligned well with the models derived using ABC, confirming the accuracy of our approach.

The rest of the paper is organized as follows. In Section \ref{SE2}, we describe our model in its different components: (i) a discrete, (ii) a stochastic discrete brain network model and (iii) the impact of Bayesian inference of the CBGTH. Section \ref{SE3} presents numerical results based on the developed stochastic discrete brain network model and ABC technique for the cortex-thalamus-basal-ganglia systems. The computational results were obtained using codes developed in C-language and SHARCNET supercomputer facilities, and the simulation results were visualized in MATLAB. Implications of these results and their importance are discussed in Section \ref{SE4}. Finally, we conclude our findings and outline future directions in Section \ref{SE5}.

\section{Methodology} \label{SE2}
This section highlights the discrete and stochastic brain network model of CBGTH. In this section, we present (a) the discrete model of the CBGTH network mediated by Laplacian terms, (b) the stochastic brain network model of the CBGTH system, giving particular attention to stochastic noises and (c) the Bayesian inference for stochastic neural network models such as brain network model of the CBGTH system. We evaluated the behaviour of stochastic noises in the brain regions such as Gpe, GPi, STN and thalamus (TH) and firing patterns under healthy and pathological states to validate the features of the CBGTH model. We then use data to examine the firing rates of the coupled neurons on each node in the brain network. Additionally, the effects of noise in the presence of DBS on TH, STN, GPi and GPe are evaluated. Finally, the impact of Bayesian inference on estimating the posterior distributions of the parameters in
a stochastic brain network model of the CBGTH has been analyzed.
\subsection{Discrete brain network model of CBGTH}
\label{Section2.2}
The network comprises nodes delineated within the brain connectome, often corresponding to established brain atlas regions. We aim to construct a model capable of capturing temporal voltage variations across different nodal points.

The brain connectome is represented as a weighted network $\mathcal G$ consisting of $V$ nodes and $E$ edges, derived from diffusion tensor imaging (DTI) and tractography techniques \cite{shaheen2022multiscale}, as adopted from the HCP dataset. The edges of this network symbolize axonal bundles within white-matter tracts. To generate a network approximation of the diffusion terms, we utilize a weighted graph Laplacian, where the weights of the weighted adjacency matrix $\textbf{W}$ are determined by the ratio of the mean fibre number $n_{ij}$ to the mean squared length $l^2_{ij}$ connecting nodes $i$ and $j$, expressed as:
\begin{equation}\label{1}
W_{ij}= \frac{n_{ij}}{l^2_{ij}}, \quad i=1,\dots, V.
\end{equation}
These weights align with the inverse length-squared dependency observed in the canonical discretization of the continuous Laplace (diffusion) operator \cite{shaheen2022multiscale}. Additionally, we define the diagonal weighted degree matrix as:
\begin{equation}\label{2}
D_{ii}=\sum_{j=1}^V W_{ij}, \quad i,j=1,\dots, V.
\end{equation}
Furthermore, the graph Laplacian $\textbf{L}$ with $(i,j)$-entry is defined as:
\begin{equation}\label{3}
L_{ij}= \rho (D_{ij}-W_{ij}), \quad i,j=1,\dots, V,
\end{equation}
where $\rho$ represents the diffusion coefficient.

\begin{figure}[h]
\begin{subfigure}{0.4\textwidth}
    \includegraphics[width=\textwidth]{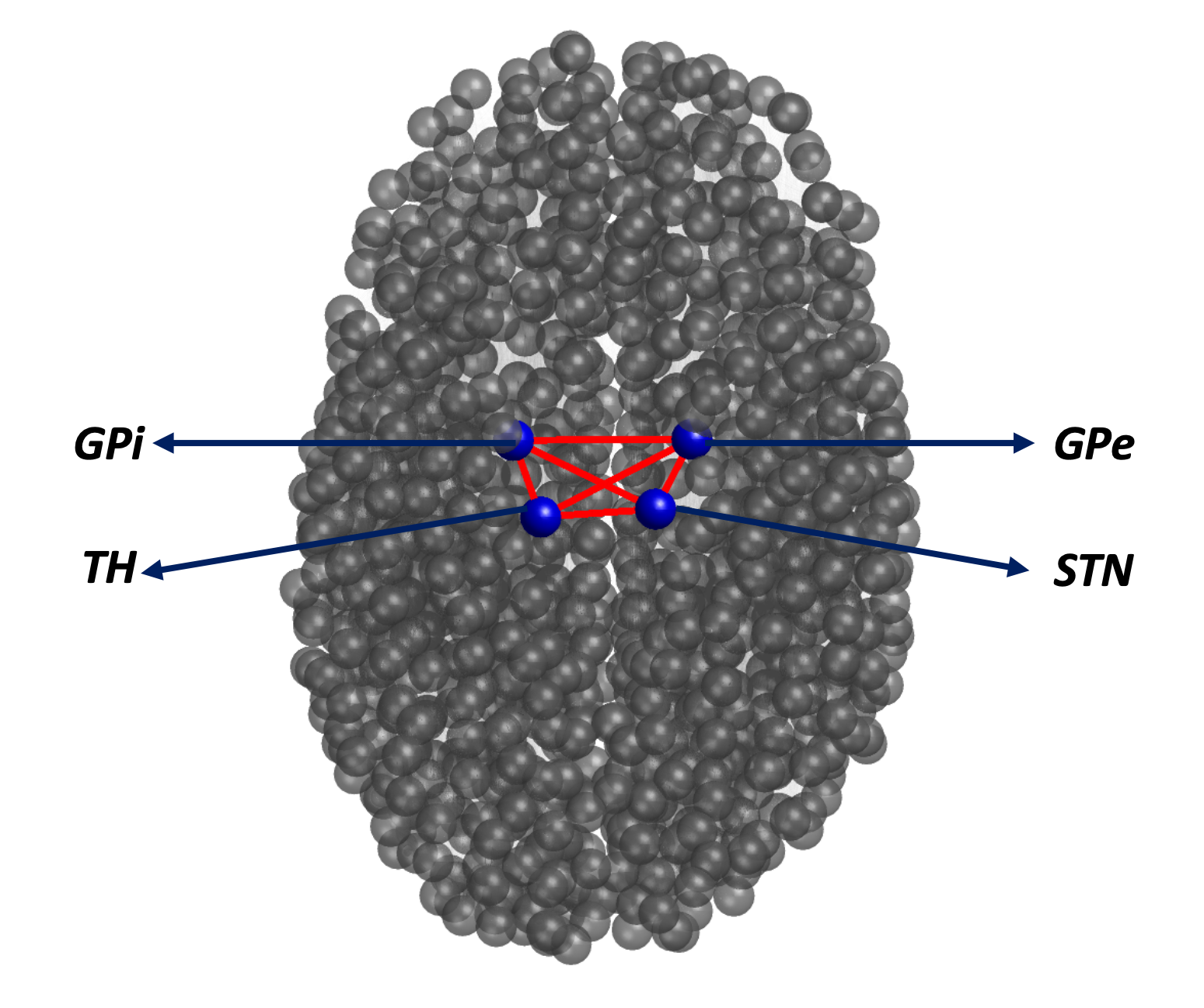}
    \caption{}

\end{subfigure}
\begin{subfigure}{0.5\textwidth}
    \includegraphics[width=\textwidth]{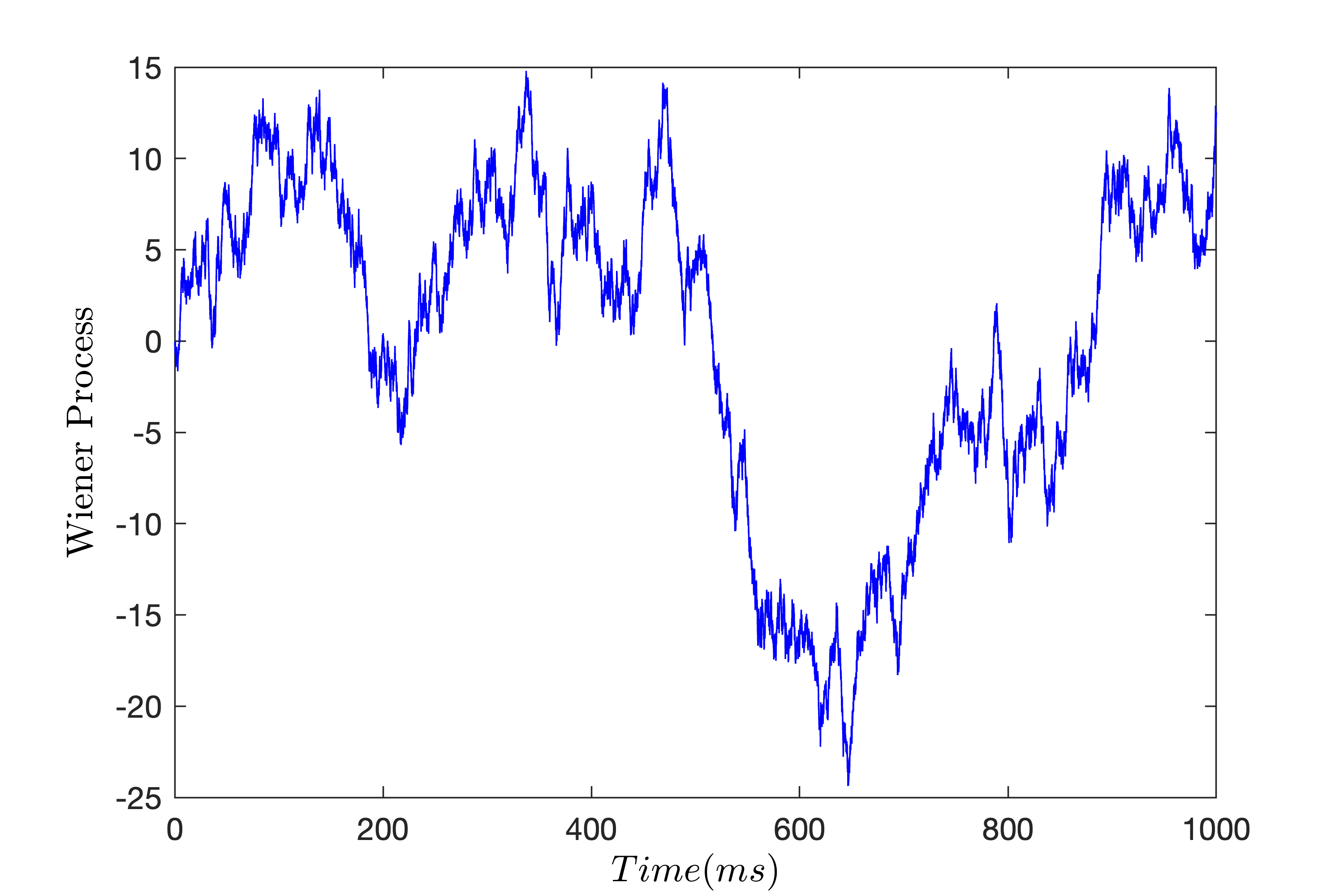}
    \caption{}

\end{subfigure}
\caption{(Color online) (a) Discrete brain network connectome in a healthy condition (left) (axial view from bottom). The four nodes are STN, GPe, GPi, and TH, and we replaced the spiking node "cortex" with the whole brain connectome (b) Stochastic noises applied to STN, GPe, GPi, and TH } 
\label{fig1}
\end{figure}
The adjacency matrix for simulations is derived from diffusion tensor magnetic resonance images obtained from $418$ healthy HCP subjects sourced from the Budapest Reference Connectome v3.0 \cite{shaheen2022multiscale}. Fig.~\ref{fig1} (a) showcases a network composed of $V = 4$ nodes and $E = 6$ edges representing brain regions like the putamen, globus pallidus, and thalamus. Each node is assumed to occupy a surface area of $1.5{\si{cm}^2}$. Each node linked with STN, GPi, GPe, and TH carries the voltage $v^{sn}, v^{gi}, v^{ge}$, and $v^{th}$, respectively. The network equations for the continuous model take the form of a system of first-order ordinary differential equations as follows:
\begin{equation}\label{smm}
    \frac{dv^{sn}}{dt} =-d_{v^{sn}}\sum_{k=1}^VL_{1k}{v_k}+\frac{1}{c_m}\bigg(-I_{Na}^{sn}-I_K^{sn}-I_L^{sn}-I_T^{sn}-I_{Ca}^{sn}-I_{ahp}^{sn}-I_{ge\rightarrow{sn}}+I_{snapp}\bigg),
\end{equation} 
\begin{eqnarray}
    \frac{dv^{gi}}{dt} =-d_{v^{gi}}\sum_{k=1}^VL_{2k}{v_k}+\frac{1}{c_m}\bigg(-I_{Na}^{gi}-I_K^{gi}-I_L^{gi}-I_T^{gi}- \\  \nonumber I_{Ca}^{gi}-I_{ahp}^{gi} -I_{sn\rightarrow{gi}}-I_{ge\rightarrow{gi}}+I_{giapp}\bigg),
\end{eqnarray}
\begin{eqnarray}
    \frac{dv^{ge}}{dt}=-d_{v^{ge}}\sum_{k=1}^VL_{3k}{v_k}+\frac{1}{c_m}\bigg(-I_{Na}^{ge}-I_K^{ge}-I_L^{ge}-  I_T^{ge}- \\  \nonumber I_{Ca}^{ge}-I_{ahp}^{ge} -I_{sn\rightarrow{ge}}-I_{ge\rightarrow{ge}}+I_{geapp}\bigg),
\end{eqnarray}
\begin{equation}\label{sum1}
    \frac{dv^{th}}{dt}
    =-d_{v^{th}}\sum_{k=1}^VL_{4k}{v_k}+\frac{1}{c_m}\bigg(-I_{Na}^{th}-I_K^{th}-I_L^{th}-I_T^{th}-I_{gi\rightarrow{th}}+I_{smc}\bigg),
\end{equation}
with non-negative initial conditions for all variables $v^{sn}$, $v^{gi}$, $v^{ge}$, and $v^{th}$. Additionally, $d_{v^{sn}}, d_{v^{gi}}, d_{v^{ge}}$, and $d_{v^{th}}$ represent the diffusion terms corresponding to each node. The weights in the weighted adjacency matrix represent the spread of transneuronal degeneration from one node to its neighbours. Next, we introduce stochastic noise into the discrete brain network model to observe its influence.

\subsection{Stochastic brain network model of CBGTH }
The integration of ML techniques with stochastic modelling in brain studies holds significant promise for advancing our understanding of neural dynamics and function \cite{wozniak2020deep}. In this section, we develop a discrete brain network model incorporating the addition of stochastic noise. The noise levels are crucial for ensuring the proper functioning of signals within the nervous system \cite{shi2023characteristic}. Studies have suggested that in computational models of neurodegenerative conditions such as PD, increased external noise levels are necessary for optimal function, reflecting the aging process and reduced plasticity \cite{liu2018noise}. Consequently, noise stimulation could be an alternative therapeutic approach for alleviating PD symptoms \cite{shi2023characteristic}. Therefore, based on the model presented in Section \ref {Section2.2}, we have added the noise terms as follows:
\begin{eqnarray}\label{sum3}
\frac{dv^{sn}}{dt} =-d_{v^{sn}}\sum_{k=1}^VL_{1k}{v_k}+\frac{1}{c_m}\bigg(-I_{Na}^{sn}-I_K^{sn}-I_L^{sn}-I_T^{sn}-I_{Ca}^{sn}-  \\ \nonumber I_{ahp}^{sn}-I_{ge\rightarrow{sn}}+I_{snapp}\bigg) +  \sigma_{1} \cdot dW_1(t),
\end{eqnarray}
\begin{eqnarray}
\frac{dv^{gi}}{dt} =-d_{v^{gi}}\sum_{k=1}^VL_{2k}{v_k}+\frac{1}{c_m}\bigg(-I_{Na}^{gi}-I_K^{gi}-I_L^{gi}-I_T^{gi}-I_{Ca}^{gi}-\\ \nonumber I_{ahp}^{gi} -I_{sn\rightarrow{gi}}-I_{ge\rightarrow{gi}}+I_{giapp}\bigg) + \sigma_{2} \cdot dW_2(t),
        \end{eqnarray}
\begin{eqnarray}
\frac{dv^{ge}}{dt} =-d_{v^{ge}}\sum_{k=1}^VL_{3k}{v_k}+\frac{1}{c_m}\bigg(-I_{Na}^{ge}-I_K^{ge}-I_L^{ge}-I_T^{ge}-I_{Ca}^{ge}-\\ \nonumber I_{ahp}^{ge} -I_{sn\rightarrow{ge}}-I_{ge\rightarrow{ge}}+I_{geapp}\bigg) + \sigma_{3} \cdot dW_3(t),
\end{eqnarray}
\begin{equation}\label{sum4}
\frac{dv^{th}}{dt} =-d_{v^{th}}\sum_{k=1}^VL_{4k}{v_k}+\frac{1}{c_m}\bigg(-I_{Na}^{th}-I_K^{th}-I_L^{th}-I_T^{th}-I_{gi\rightarrow{th}}+I_{smc}\bigg) +  \sigma_{4} \cdot dW_4(t),
\end{equation}
where \( dW_i(t) \) represents the increment of the Wiener process \( W_i(t) \) and \( \sigma_i \) are the scaling factors (representing the intensity of the noise) for each equation. When numerically integrating these stochastic differential equations, we generated increments of the Wiener process at each time step \( dt \) to represent the stochastic component using the Euler–Maruyama method. Incorporating noise into the CBGTH system within a discrete brain network model provides valuable insights into how the brain functions \cite{liu2018noise}. Fig.~\ref{fig1} (b) showcases a Wiener process or stochastic noises added into the CBGTH system. Since noise is present throughout various neural processes, from perceiving sensory signals to generating motor responses, it profoundly affects neuronal dynamics. Therefore, understanding the impact of noise is crucial for comprehending the brain's behaviour \cite{charalambous2023natural}. The significance of this impact will be explored further in the following Section \ref{SE3}. Moreover, the DBS current is added to the spatio-temporal model to the membrane potential equations of STN as follows: 
\begin{eqnarray}
\label{dbs}
    \frac{dv^{sn}}{dt} =-d_{v^{sn}}\sum_{k=1}^VL_{1k}{v_k}+\frac{1}{c_m}\bigg(-I_{Na}^{sn}-I_K^{sn}-I_L^{sn}-I_T^{sn}-I_{Ca}^{sn}-  \\ \nonumber I_{ahp}^{sn}-I_{ge\rightarrow{sn}}+I_{snapp}+I_{DBS}\bigg) + \sigma_{1} \cdot dW_1(t),
\end{eqnarray}
where $c_m=1\mu F/{cm}^2$ and $ I_{DBS}$ is adopted from  \cite{shaheen2022multiscale}. According to Eq. (\ref{dbs}), the DBS electrode has been applied to the STN node in the discrete brain network connectome. The relevant parameters are given in Table \ref{tab:1} (the other relevant parameters are adopted from \cite{shaheen2022multiscale}, ($pd$ is a parameter, and $pd=0$ indicates that the network is in healthy states, while $pd=1$ shows that the network is in Parkinsonian states). The $--$ represents no connection to neurons.
\begin{table}
\caption{Parameter set for the CBGTH network \cite{shaheen2022multiscale}.}
\centering
\footnotesize
\label{tab:1}       
%
%
\begin{tabular}{p{1cm}p{3.5cm}p{3.5cm}p{3.5cm}}
\hline\noalign{\smallskip}
& STN neuron & GPe/GPi neuron & TH neuron \\
\hline\noalign{\smallskip}
$I_{Ca}$  &$2(c^2)(v-140)$  &$0.15(s_{\infty}(v))^2(v-120)$  &$--$\\
 $I_{ahp}$  &$20(v+80)\newline(w/(w+15))$  &$10(v+80)(w/(w+10))$ &$--$\\
 $I_{ge\rightarrow{sn}}$&$0.5S_{ge\rightarrow{sn}}(v+85)$  & $--$&$--$\\
  $I_{ge\rightarrow{ge}}$& $--$ &$0.5S{ge\rightarrow{ge}}(v+85)$ &$--$\\
  $I_{ge\rightarrow{gi}}$& $--$ &$0.5S_{ge\rightarrow{gi}}(v+85)$ &$--$\\
  $I_{sn\rightarrow{ge}}$& $--$ &$0.15S_{sn\rightarrow{ge}}v$ &$--$\\
  $I_{sn\rightarrow{gi}}$& $--$ &$0.15S_{sn\rightarrow{gi}}v$ &$--$\\
   $I_{gi\rightarrow{th}}$& $--$ &$--$ &$0.112S_{gi\rightarrow{th}}(v+85)$\\
  $I_{snapp}$&$33-10pd$  &$--$&$--$\\
    $I_{giapp}$& &$22-6pd$ &$--$\\

   $I_{geapp}$&  &$21-13pd+(-1.5)$ &$--$\\
\end{tabular}
\end{table}
\subsection{Bayesian Inference for Stochastic Neural Network Models}
In this section, we apply Bayesian inference to estimate the posterior distributions of the parameters in a system of stochastic differential equations (i.e., Eqs. \ref{sum3}-\ref{sum4}) governing the membrane potentials of different brain regions. Since \( \sigma_i \cdot dW_i(t) \) introduces stochastic noise into the system, it represents the random fluctuations or uncertainty in the dynamics of the model. The goal is to use Bayesian inference to estimate the noise parameters \( \sigma_1, \sigma_2, \sigma_3, \sigma_4 \), which represent the intensity of the stochastic perturbations.

We aim to estimate the posterior distributions of the parameters \( \theta = (\sigma_1, \sigma_2, \sigma_3, \sigma_4) \) given the observed data \( y_{\text{obs}} \) (e.g., recorded membrane potentials). The Bayesian inference framework allows us to combine prior knowledge about the parameters with the likelihood of observing the data under a given set of parameters. Using Bayes' theorem, the posterior distribution is given by:
\begin{equation}
p(\theta | y_{\text{obs}}) = \frac{p(y_{\text{obs}} | \theta) \cdot p(\theta)}{p(y_{\text{obs}})},
\end{equation}
where \( p(y_{\text{obs}} | \theta) \) is the likelihood function, \( p(\theta) \) is the prior distribution, and \( p(y_{\text{obs}}) \) is the marginal likelihood (which can be treated as a normalizing constant). We assume normal prior distributions for the noise parameters \( \sigma_1, \sigma_2, \sigma_3, \sigma_4 \), with mean 0 and standard deviation 1:
\begin{equation}
\sigma_i \sim \mathcal{N}(0, 1), \quad \text{for } i = 1, 2, 3, 4.
\end{equation}
The likelihood function for the observed data \( y_{\text{obs}} \) given the parameters \( \theta \) is defined as:
\begin{equation}
p(y_{\text{obs}} | \theta) = \exp \left( -\frac{1}{2} \sum_{i=1}^{N} \left( \frac{y_{\text{obs}, i} - y_{\text{sim}, i}(\theta)}{\sigma_{\text{obs}}} \right)^2 \right),
\end{equation}
where \( y_{\text{sim}}(\theta) \) represents the data generated by the model using the parameters \( \theta \), and \( \sigma_{\text{obs}} \) is the observational noise and \( \theta_0 = (\sigma_1^0, \sigma_2^0, \sigma_3^0, \sigma_4^0) \)$=[0.5, 0.5, 0.5, 0.5]$. To estimate the posterior distributions of the parameters, we use the Markov Chain Monte Carlo (MCMC) method. MCMC generates samples from the posterior distribution by constructing a Markov chain that has the desired posterior distribution as its equilibrium distribution. In this case, we employ the Metropolis-Hastings algorithm, which proposes new parameter values and accepts or rejects them based on the posterior probability. The MCMC algorithm proceeds as follows:

\begin{enumerate}
    \item Initialize the parameter values \( \theta_0 = (\sigma_1^0, \sigma_2^0, \sigma_3^0, \sigma_4^0) \)
    \item For each iteration \( i \):
    \begin{enumerate}
        \item Propose new values \( \theta' = (\sigma_1', \sigma_2', \sigma_3', \sigma_4') \) by adding small random noise to the current values.
        \item Simulate data \( y_{\text{sim}} \) using the proposed parameter values.
        \item Compute the likelihood of the observed data given the proposed parameters.
        \item Accept the new parameter values with probability:
        \begin{equation}
        \alpha = \min \left( 1, \frac{p(y_{\text{obs}} | \theta') p(\theta')}{p(y_{\text{obs}} | \theta) p(\theta)} \right),
        \end{equation}
        otherwise, retain the current values.
    \end{enumerate}
\end{enumerate}
Finally, after $10000$ iterations, the accepted parameter values form an approximation to the posterior distribution, and it emphasizes the benefits of Bayesian inference in providing uncertainty quantification for the estimated parameters. The parameters representing the intensity of this noise are key factors influencing the behaviour of each brain region. Through Bayesian inference, we estimate the posterior distribution of these noise parameters, providing insight into how uncertainty and variability impact neural dynamics, as given next in Section \ref{SE3}. This approach allows us to quantify the influence of noise on brain activity and provide a deeper understanding of how stochastic processes contribute to normal function and dysfunction in neural systems. Therefore, by incorporating the posterior distributions of the noise terms, we can better understand the variability in brain activity across these regions, such as STN, GPE, GPi, and TH and how this variability might affect processes like signal transmission, synchronization, and information processing within the brain's neural circuits. Importantly, \(\sigma_1, \sigma_2, \sigma_3, \sigma_4\) represent the standard deviations of the membrane potentials of \(\sigma_{v^{sn}}, \sigma_{v^{gpe}}, \sigma_{v^{gpi}}, \sigma_{v^{th}}\).

\section{Results} \label{SE3}
In this section, we will investigate how stochastic noise impacts the CBGTH system in healthy and PD brain states by integrating ML and stochastic modelling with connectomic data. Additionally, we investigate the ABC technique to estimate the posterior distributions of the parameters in
a CBGTH system.  

Importantly, noise introduces stochastic fluctuations into the brain network, affecting the timing and reliability of neural signal transmission \cite{shaheen2022deep,zheng2023noise}. In the context of the basal ganglia-thalamocortical circuit, where precise timing is crucial for motor control and cognitive processes, the impact of noise may lead to alterations in information processing and integration \cite{shaheen2022deep}. Moreover, neural noise, originating from various sources such as sensory input, cellular processes, and electrical activity, significantly influences the functioning of the nervous system. While it can hinder information processing, it also contributes to brain function by shaping functional networks, enhancing synchronization, and impacting task performance \cite{zheng2023noise}. The brain's dynamics, characterized by subject-specific parameters and diverse outputs, make it a noisy dynamical system. Recent research indicates that noninvasive brain stimulation can alter the signal-noise relationship, but the precise relationship between noise amplitude and the global effects of local stimulation remains uncertain \cite{shaheen2022deep}. 

\begin{figure}[h]

    \includegraphics[width=\textwidth]{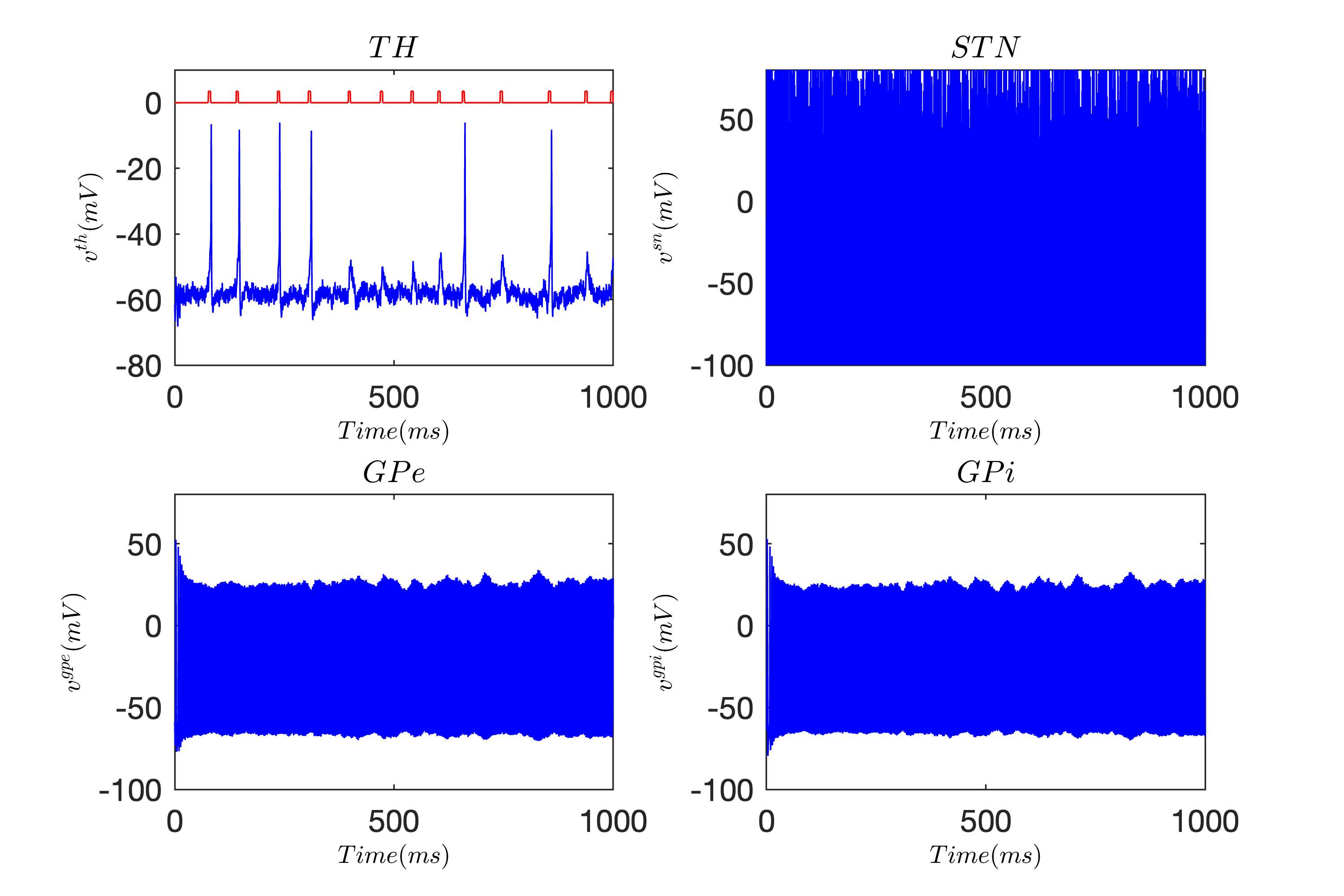}

\caption{(Color online) The effect of stochastic noise on the membrane voltages of the discrete brain network's TH, STN, GPe and GPi neurons in a healthy state. The red pulse trains in the top right panel denote SMC signal.} \label{fig2}
\end{figure}
The impact of stochastic noise on the selected regions, namely the  TH, STN, GPi, and GPe, in the healthy brain is illustrated in Fig.~\ref{fig2}. It is observed that burst oscillations occur across all brain regions, particularly in the thalamus and subthalamic nucleus, where the oscillations persistently burst. Consequently, this heightened neural activity exacerbates the healthy state of the brain. As a result, it impairs membrane potential and disrupts the normal functioning of neurons. These findings underscore the significant adverse effects of stochastic noise on the brain and its constituent regions, potentially leading to the development or exacerbation of brain injury \cite {touboul2020noise,staffaroni2019longitudinal}.

\begin{figure}[h]
    \includegraphics[width=\textwidth]{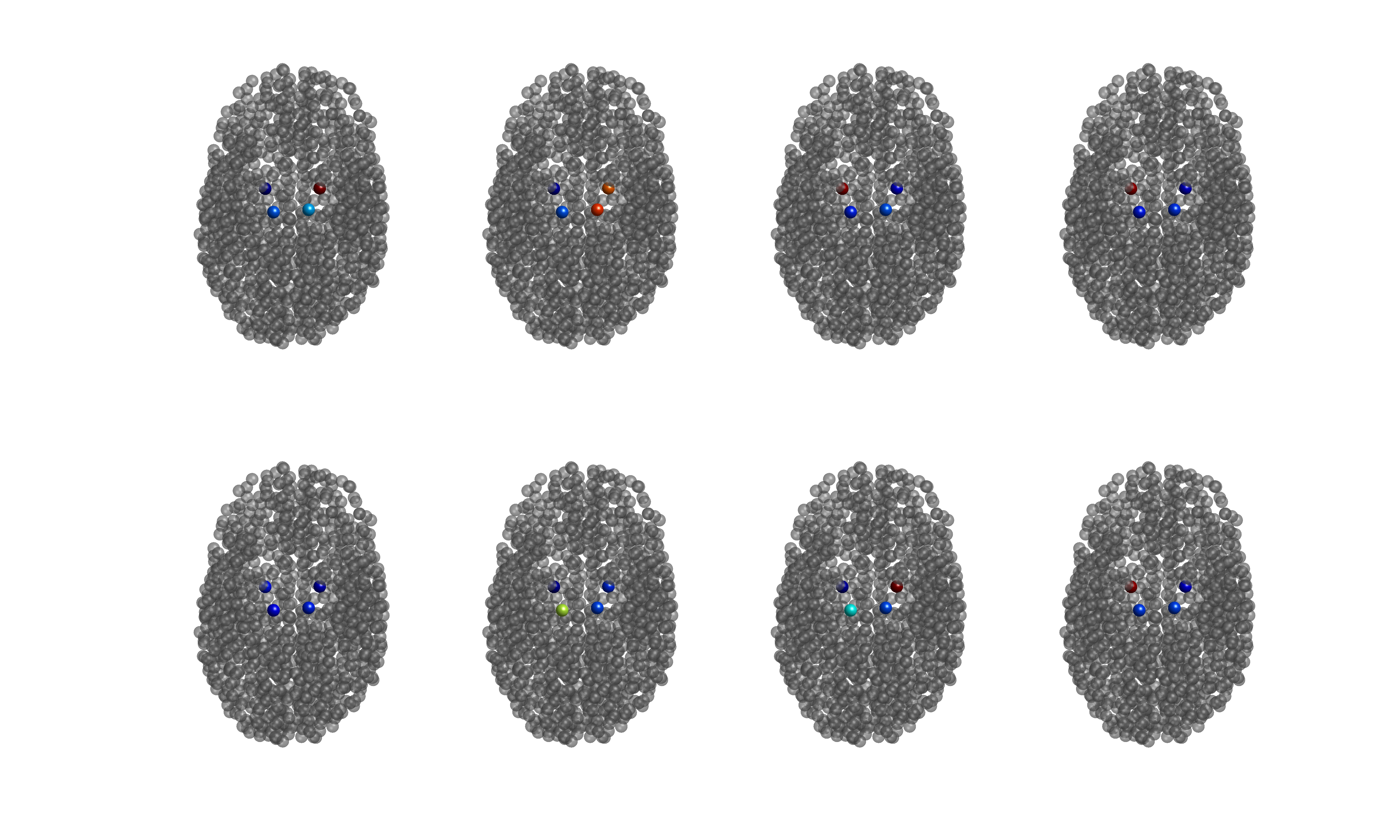}

\caption{(Color online) The effect of stochastic noise on the membrane voltage distributions of TH (top right), STN (top left), GPe (bottom right) and GPi (bottom left) neurons in the brain over time in Parkinson's
state (axial views from below). Top panel (left to right): $t=354.56\si{ms}, 356.8\si{ms}, 356.23\si{ms}, 370\si{ms}$, and for the bottom panel (left to right): $t=374.67\si{ms}, 383.2\si{ms}, 385.05\si{ms}, 385.57\si{ms}$.} \label{fig3}
\end{figure}
In the Parkinsonian state, the membrane voltages of key neuronal populations, including STN, GPi, GPe, and TH neurons within a discrete brain network, exhibit dynamic fluctuations over time, as depicted in Fig.~\ref{fig3}. The initial equilibrium has been set to $-65mV$; these neurons display varying voltage concentrations due to stochastic effects and diffusion processes. The color scale of voltage concentrations in Fig.~\ref{fig3} is plotted using MATLAB jet colormap. The recorded voltages at specific time points, such as $t = 354.56ms, 356.8ms, 356.23ms$, $370ms,$ $ 374.67ms, 383.2ms, 385.05ms$, and $385.57ms$, reveal temporal changes in neuronal activity. Notably, certain neuronal populations exhibit elevated voltages relative to others at different time points, as indicated by the color nodes. For instance, at $t = 356.23ms$, the voltage of TH neurons surpasses that of other neurons. In contrast, at $t = 356.8ms$, the GPe and TH neurons exhibit higher voltage concentrations within the CBTH circuitry in the Parkinsonian state. These voltage dynamics underscore the intricate interplay of stochastic noise and diffusion processes in shaping neuronal activity patterns associated with PD \cite{shaheen2022multiscale}.

In Fig.~\ref{fig3}, stochastic noise is crucial in modulating the membrane voltage distributions of key neuronal populations implicated in PD. Over time, stochastic fluctuations in membrane potentials within these neural networks can exacerbate pathological activity patterns in the Parkinsonian state. In the TH neurons involved in dopamine production, stochastic noise may contribute to the dysregulation of dopamine levels characteristic of Parkinson's. Similarly, in the STN, known for its involvement in motor control, stochastic noise might amplify aberrant firing patterns associated with movement dysfunction. Meanwhile, within the GPe and GPi, integral components of the basal ganglia circuitry, stochastic fluctuations could disrupt the delicate balance of inhibitory signalling, further exacerbating motor symptoms \cite{wozniak2020deep,liu2018noise}. These stochastic influences underscore the complexity of Parkinson's pathophysiology and highlight the importance of understanding noise modulation within neural circuits for developing effective therapeutic interventions \cite{touboul2020noise}.

\begin{figure}[h]
    \includegraphics[width=\textwidth]{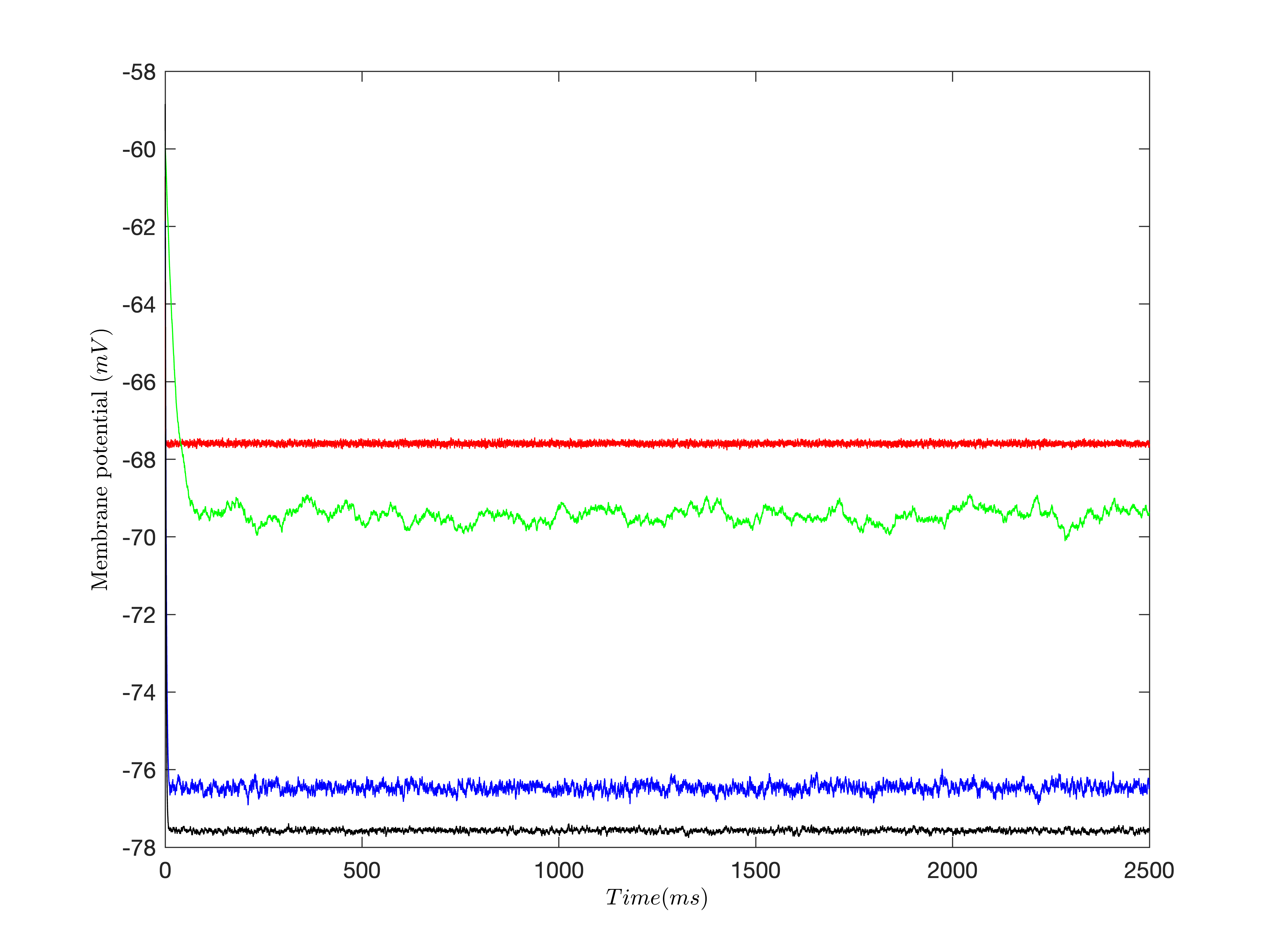}

\caption{(Color online) The membrane potential fluctuations over time of TH (red curve), STN (green curve), GPe (blue curve) and GPi (black curve) neurons in the brain over time in Parkinson's
state under noisy conditions.} \label{fig3a}
\end{figure}
Fig.~\ref{fig3a} depicts membrane potential fluctuations over time in four different brain regions, including STN, GPi, GPe, and TH neurons within a discrete brain network in Parkinson's state under noisy conditions. All regions seem to experience an initial rapid depolarization or change in membrane potential followed by stabilization \cite{volgushev1998modification}. This initial drop indicates the response of the neurons to an input or perturbation, which might be electrical stimulation or synaptic input, followed by settling into a steady-state potential \cite{gutkin2005phase}. For instance, TH shows the least fluctuation around a steady value (approximately $-68 mV$). This suggests that the thalamus region has a higher resistance to noise or exhibits more stable dynamics compared to other regions. Similarly, STN displays moderate noise fluctuations around its resting membrane potential (around $-70 mV$). The noise causes visible variability, but the STN remains relatively stable over time. Moreover, GPe exhibits slightly larger noise-induced fluctuations than the STN, with its resting potential slightly lower (around $-75 mV$). The GPe seems more sensitive to noise compared to the TH and STN. Finally, GPi shows the lowest membrane potential (around $-77 mV$) with small fluctuations, suggesting this region also has stable membrane dynamics, similar to the thalamus. This difference could be significant when considering how these regions contribute to network dynamics in disease conditions such as Parkinson's or other movement disorders \cite{kim2017abnormal}. These fluctuations in the STN and GPe could imply that they play a more active role in relaying noisy inputs, possibly affecting downstream targets.

\begin{figure}[h]
    \includegraphics[width=\textwidth]{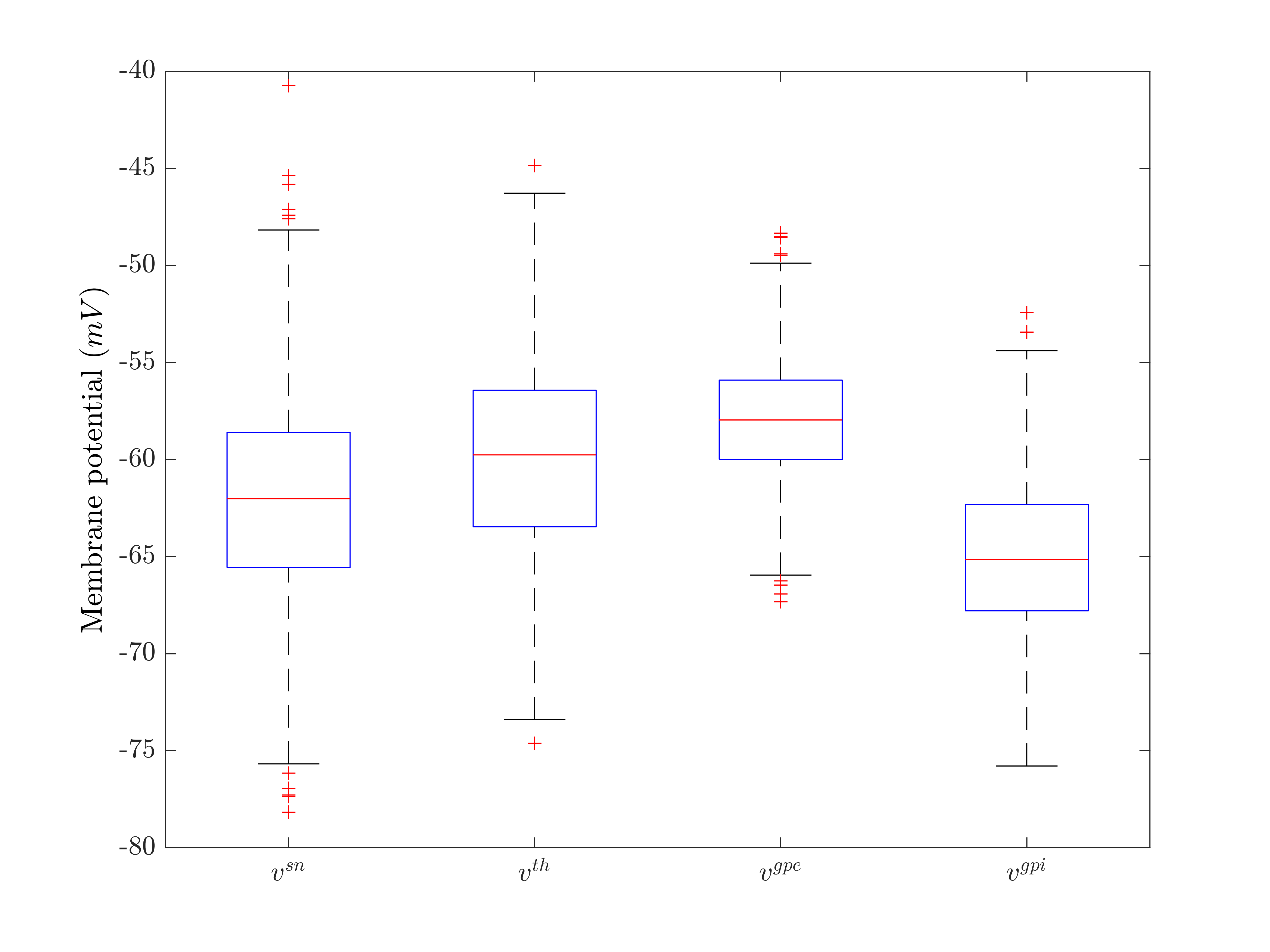}

\caption{(Color online)The distribution of membrane potential for each brain region: STN, TH, GPe and GPi based on the brain network data with noise. } \label{fig3ab}
\end{figure}
In Fig.~\ref{fig3ab}, the box plots represent the distribution of membrane potential for each of the brain regions: STN, TH, GPe and GPi based on the brain network data with noise. Each box plot shows the median value of the membrane potential (represented by the red line inside the box) for the four regions. The height of each box represents the interquartile (IQR), indicating the range between the $25^{th}$ and $75^{th}$ percentiles of the membrane potentials. The whiskers extend to $1.5$ times the IQR, representing the range of the bulk of the data, while the red crosses indicate outliers. The box plots represent the distribution of membrane potentials for the STN, TH, GPe and GPi under noisy conditions, revealing key insights into the noise sensitivity across these brain regions. The median membrane potentials are relatively similar, suggesting comparable average behaviours. At the same time, in  Fig.~\ref{fig3ab}, the IQR ranges show that the STN, TH, and GPe experience moderate variability, and the GPi has a narrower IQR, indicating more stability against noise. Outliers, particularly in the STN and GPe, indicate rare instances of extreme fluctuations, with the STN showing greater sensitivity to noise, which aligns with its role in propagating excitatory inputs within the basal ganglia \cite{deffains2016subthalamic}. The GPi’s stability may suggest its role as a stabilizer within the network, while the higher variability in the STN and thalamus highlights their susceptibility to noise and their potential involvement in pathological oscillations observed in movement disorders like PD. Overall, these box plots visualize how noise impacts the membrane potential dynamics, supporting the broader discussion of network stability and region-specific responses to external perturbations in both healthy and diseased states.

\begin{figure}[h]
    \includegraphics[width=\textwidth]{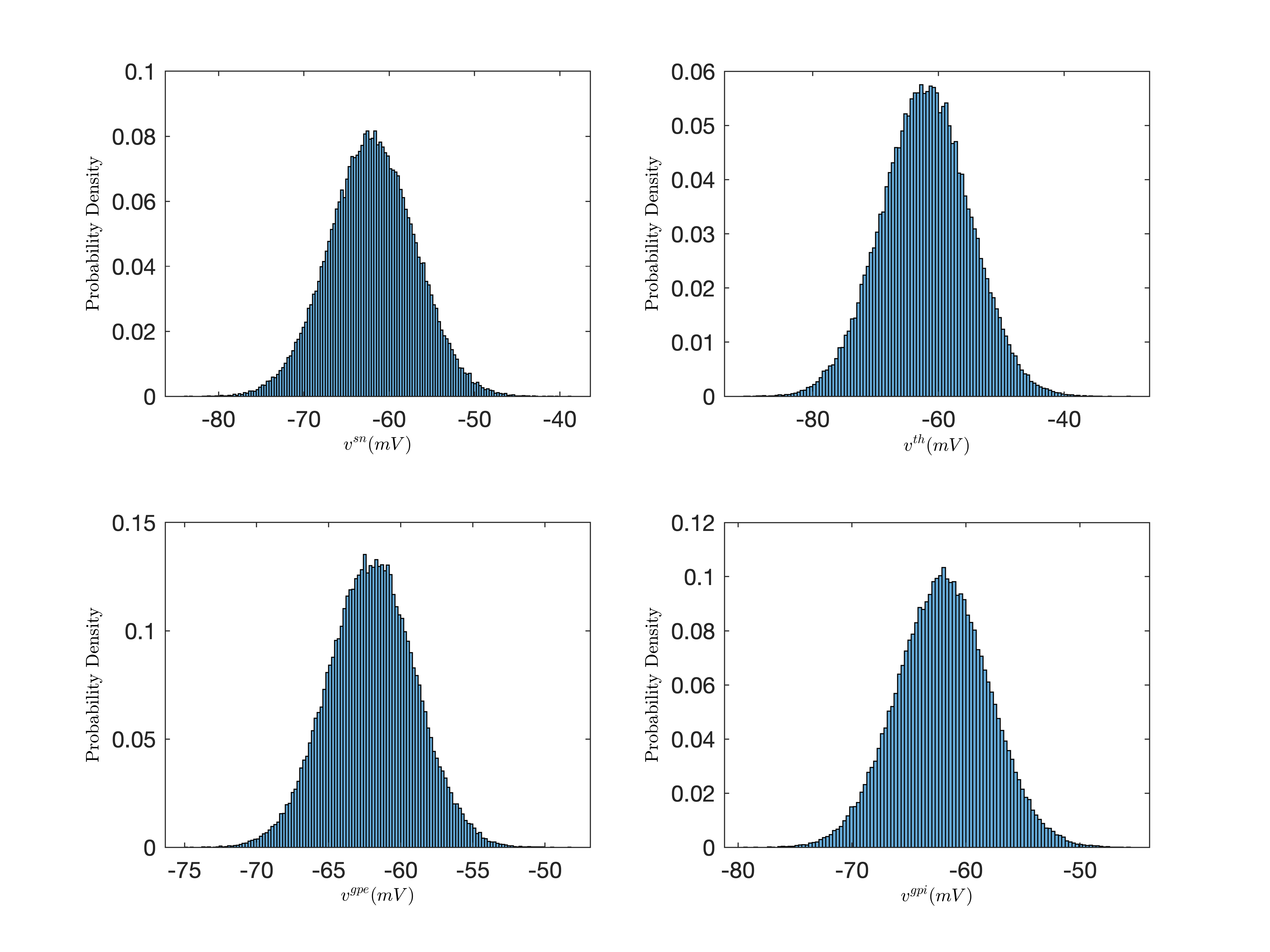}
\caption{(Color online) The probability density of membrane potentials for the TH, STN, GPe and GPi based on brain network data under noisy conditions.} \label{fig3ab1}
\end{figure}
The histograms in Fig.~\ref{fig3ab1} depict the probability density of membrane potentials for the STN, TH, GPe and GPi based on brain network data under noisy conditions. Each histogram illustrates a bell-shaped curve that closely resembles a normal distribution, indicating that the membrane potential fluctuations in these regions are Gaussian-like in nature. This supports the assumption that noise in these regions can be modelled using a stochastic framework that aligns with Gaussian distributions. Statistically, in Fig.~\ref{fig3ab1}, the data's fit to a normal distribution suggests that the stochastic model given in (Eqs. \ref{sum3}-\ref{sum4}) is appropriate for describing membrane potential dynamics under noisy conditions. The fact that all four regions exhibit Gaussian-like behaviour suggests that the stochastic processes driving these fluctuations are similar across regions, supporting the hypothesis that noise plays a significant and predictable role in brain network dynamics. In the basal ganglia, the balance of excitatory and inhibitory signals between these regions is critical for smooth and coordinated movement. The STN, for instance, provides excitatory input to both the GPi and GPe, and disruptions in its activity (such as increased noise or abnormal oscillations) are associated with hyperactivity in the GPi, leading to motor dysfunctions like bradykinesia or tremors in Parkinson’s disease. The histograms suggest that, under noisy conditions, the membrane potential distributions remain relatively stable, which may reflect the homeostatic mechanisms that help maintain normal basal ganglia function. However, deviations in the Gaussian-like patterns, particularly the presence of outliers or shifts in the mean membrane potential, could indicate pathological changes in neural excitability, such as those observed in NDDs.

\begin{figure}[h]
    \includegraphics[width=\textwidth]{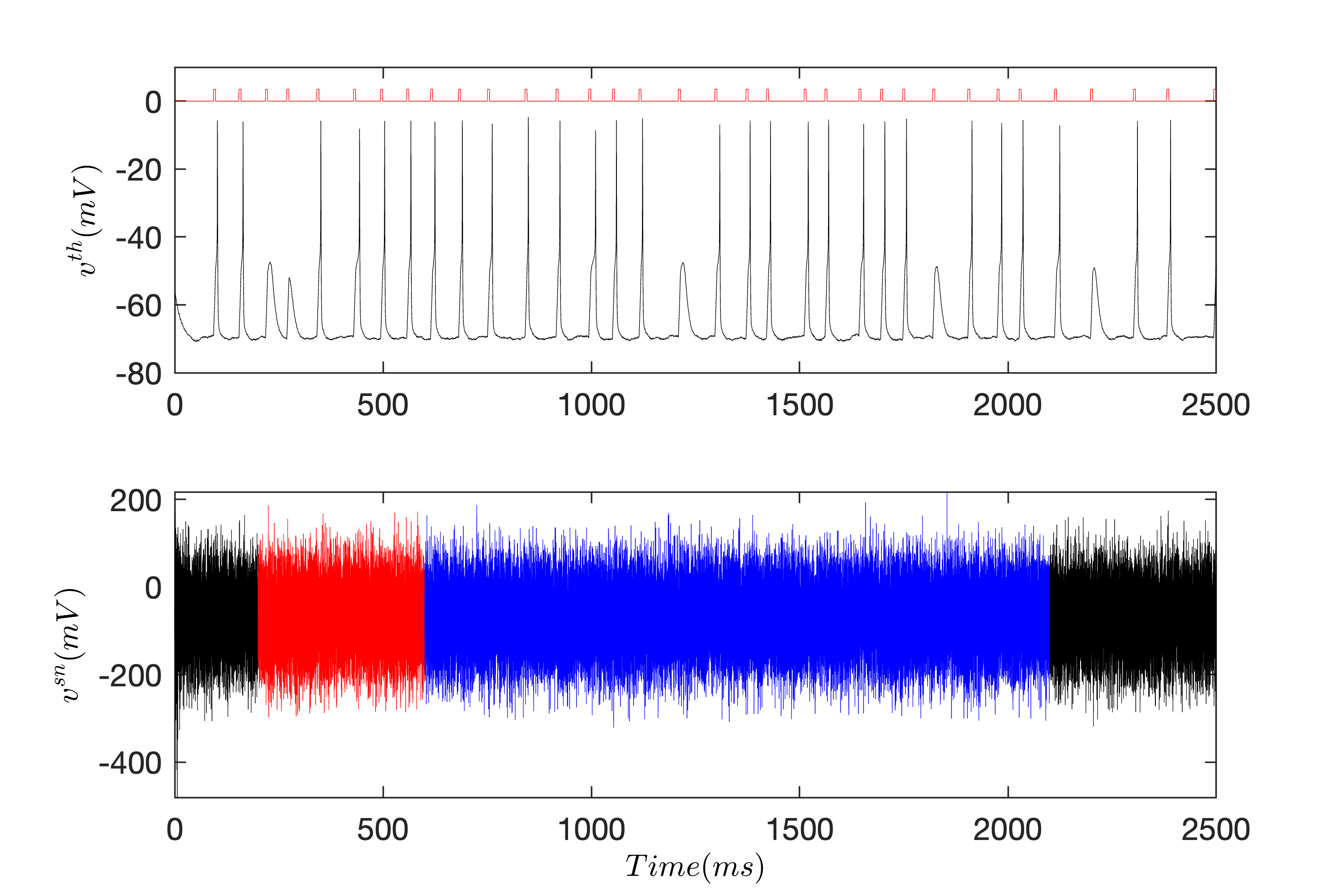}

\caption{(Color online) Effect of stochastic noise on membrane voltages of the TH (top) and the STN (bottom) neurons of discrete brain network in the
Parkinson's state (black color). The effects of open-loop DBS on the STN neurons are presented in red color (bottom). However,
the blue color shows a healthy state after the DBS is applied to the PD state. The red pulse trains in the top panel denote the SMC
signal.} \label{fig4}
\end{figure}
Next, the DBS has been applied to STN neurons in the PD state of the brain. The application of DBS to STN  neurons in the Parkinsonian state of the brain often results in a temporary restoration of healthy neural activity within the basal ganglia circuitry, leading to symptom alleviation in PD patients. As depicted in Fig.~\ref{fig4}, we applied the DBS in a spatio-temporal domain for a smaller amount of time. It is interesting to know that in the presence of a diffusion operator, neurons maintained a healthy state for a sufficient time after the DBS had been applied. We see the healthy state of STN neurons in blue color, as shown in Fig.~\ref{fig4}. However, despite the therapeutic benefits of DBS, the long-term maintenance of a healthy state remains challenging. This is evident in the observed disturbances in thalamic activity characterized by bursts of oscillations and fluctuating membrane potentials, as shown in Fig.~\ref{fig4}. Even with the presence of DBS, stochastic noise and diffusion processes continue to exert adverse effects on neural activity within the brain. Stochastic noise, arising from random fluctuations in ion channels and synaptic activity, can disrupt the finely tuned balance of excitation and inhibition within neural networks \cite{staffaroni2019longitudinal,liang2024excitation}. Additionally, diffusion processes, which govern the spread of neurotransmitters and other signalling molecules, can lead to spatial and temporal variations in neuronal activity.

In Fig.~\ref{fig4}, the persistence of disturbances in thalamic activity despite DBS suggests that stochastic noise and diffusion processes may interact with the therapeutic intervention, leading to unintended consequences. These adverse effects underscore the complexity of neural dynamics in PD and highlight the need for further research to better understand and mitigate the impact of stochastic noise and diffusion on DBS efficacy \cite{seguin2023communication,carron2013closing}. Additionally, advancements in DBS technology and optimization of stimulation parameters may help minimize these adverse effects and improve long-term therapeutic outcomes for Parkinson's patients.
\begin{figure}[h]
    \includegraphics[width=\textwidth]{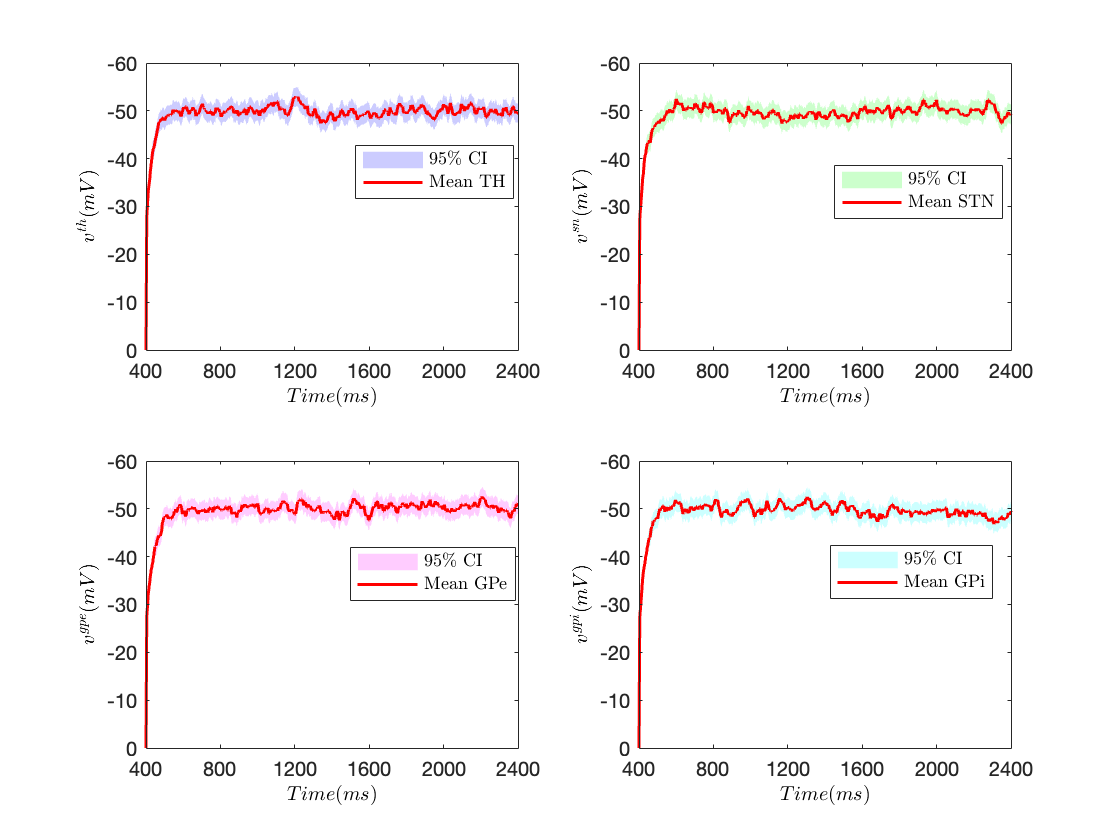}

\caption{(Color online) The provided plots display the membrane potential response for the TH, STN, GPi and GPe under the influence of DBS. Each plot shows the mean membrane potential (red line) along with the $95\%$ confidence interval (shaded area) for each region over time.} \label{fig4A}
\end{figure}
Importantly, Fig.~\ref{fig4A} represents the membrane potential response when the DBS has been applied to TH, STN, GPi and GPe neurons in the Parkinsonian state. Across all four regions, the membrane potentials tend to stabilize around a consistent value after an initial transient response to the applied DBS. The confidence intervals are relatively narrow after the initial period, showing a reduced variability in membrane potential, indicating that DBS has successfully driven the neurons towards a more stable, healthy state. This stabilization is an essential hallmark of how DBS can mitigate the pathological activity associated with disorders like PD by restoring the normal functioning of these regions. The thalamus and STN show a rapid return to a stable membrane potential after the application of DBS, as seen in the narrow confidence intervals and the smoothness of the curves. This reflects the effectiveness of DBS in normalizing excitatory-inhibitory balance in these regions, which are critically involved in motor control. Specifically, DBS likely reduces the aberrant firing patterns in the STN, a common pathological feature in PD. Similarly, the GPe and GPi exhibit a rapid return to stable membrane potentials after the application of DBS. The relatively small confidence intervals in these regions further suggest that DBS has effectively dampened the noise and reduced variability, resulting in stable network dynamics. The GPi, being the major output nucleus of the basal ganglia, is crucial for regulating motor activity, and the stabilization observed here under DBS is particularly significant for the therapeutic modulation of motor symptoms.

The confidence intervals in all regions are indicative of the membrane potential returning to within the healthy physiological range. This restoration of membrane potential stability is crucial for proper motor function, as excessive variability or deviation from the healthy range is often associated with abnormal oscillations and motor dysfunction, such as tremors or bradykinesia in PD \cite{sumarac2024clinico}. The results presented in Fig.~\ref{fig4A} highlight the effectiveness of DBS in modulating membrane potentials across the basal ganglia network and returning the system to a healthy state. The narrow confidence intervals suggest that DBS not only stabilizes the membrane potential but also reduces the variability and noise-induced fluctuations, which are commonly associated with pathological conditions. By showing that all four regions tend to a stable, healthy state with DBS, this supports the therapeutic potential of DBS in mitigating the disruptive network dynamics found in neurodegenerative diseases, particularly those involving the basal ganglia, such as Parkinson’s.

\begin{figure}[h]
    \includegraphics[width=\textwidth]{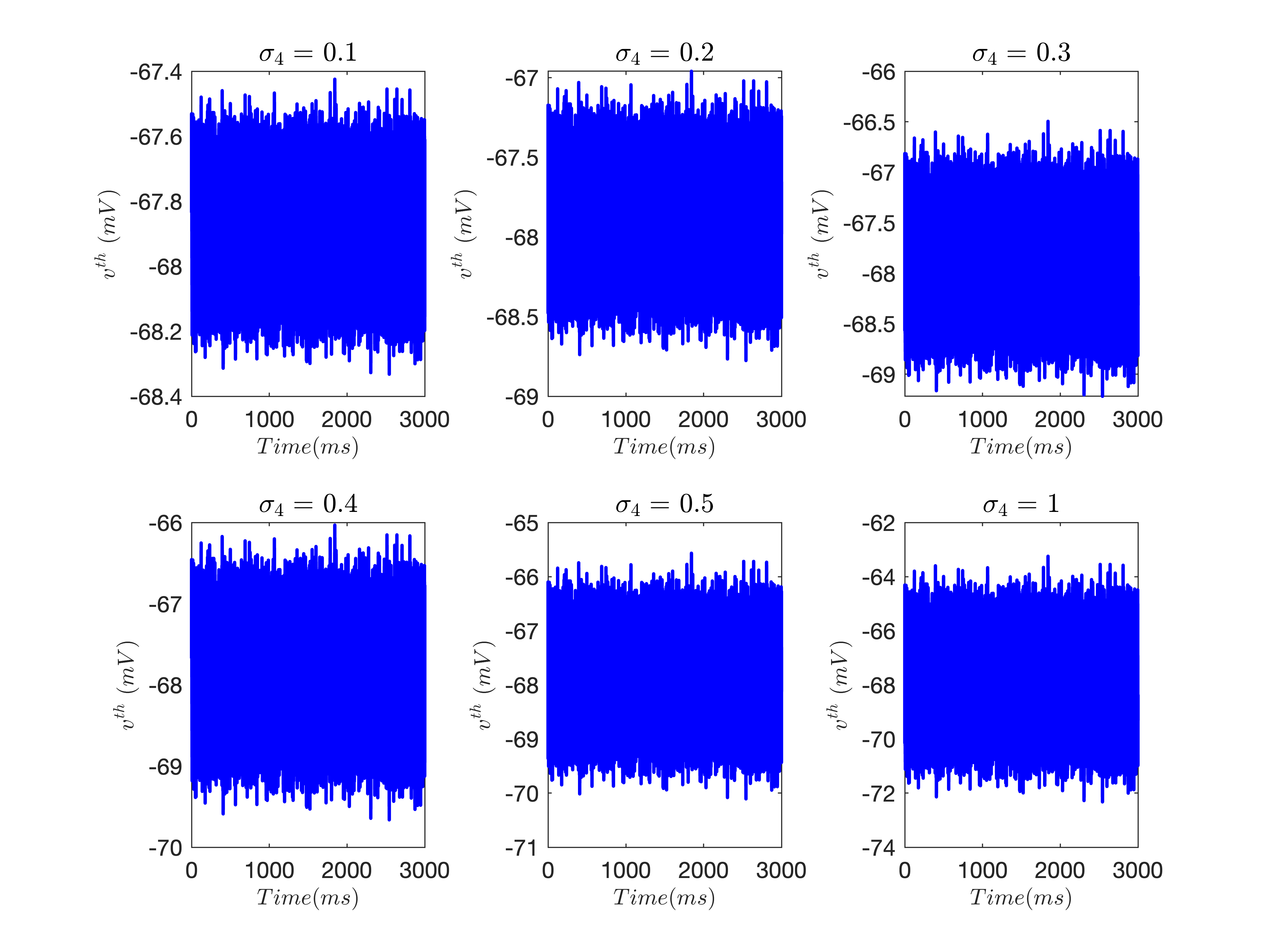}

\caption{Color online) The effect of stochastic noise on pathological activity patterns within the thalamus.} \label{fig5}
\end{figure}
Stochastic noise can significantly impact thalamic activity, disrupting its normal functioning. As depicted in Fig.~\ref{fig5}, the thalamus may exhibit erratic fluctuations in membrane potentials and firing patterns in the presence of stochastic noise. This can lead to disturbances in sensory processing, motor control, and cognitive functions that rely on thalamic signalling. Moreover, Fig.(~\ref{fig2}-\ref{fig5}) were plotted using $\sigma_{1}=0.1,\sigma_{2}=0.4,\sigma_{3}=0.4,\sigma_{4}=0.5$. As seen in Fig.~\ref{fig5}, we observed that stochastic noise tends to drive the membrane potential of thalamic neurons towards the PD state. The fluctuations in the membrane potential exhibit low-frequency oscillations. These oscillations appear to have a regular pattern but are modulated by the stochastic noise added to the system. The frequency and amplitude of these oscillations may vary depending on the system's parameters and the noise level. The stochastic noise introduced in the system causes the membrane potential to fluctuate randomly around a mean value. As the noise level ($\sigma_4$) increases, the amplitude of the fluctuations also increases. This suggests that noise can significantly influence the dynamics of the system. Incorporating stochastic noise in the thalamic membrane potential exacerbates the pathological activity patterns associated with PD. Conversely, when the noise is absent, the membrane potentials tend towards a healthier state, particularly in the presence of DBS. These observations underscore the critical role of stochastic noise in modulating thalamic activity, thereby influencing the balance between pathological and healthy states in neurological disorders such as PD.

\begin{figure}[h]
    \includegraphics[width=\textwidth]{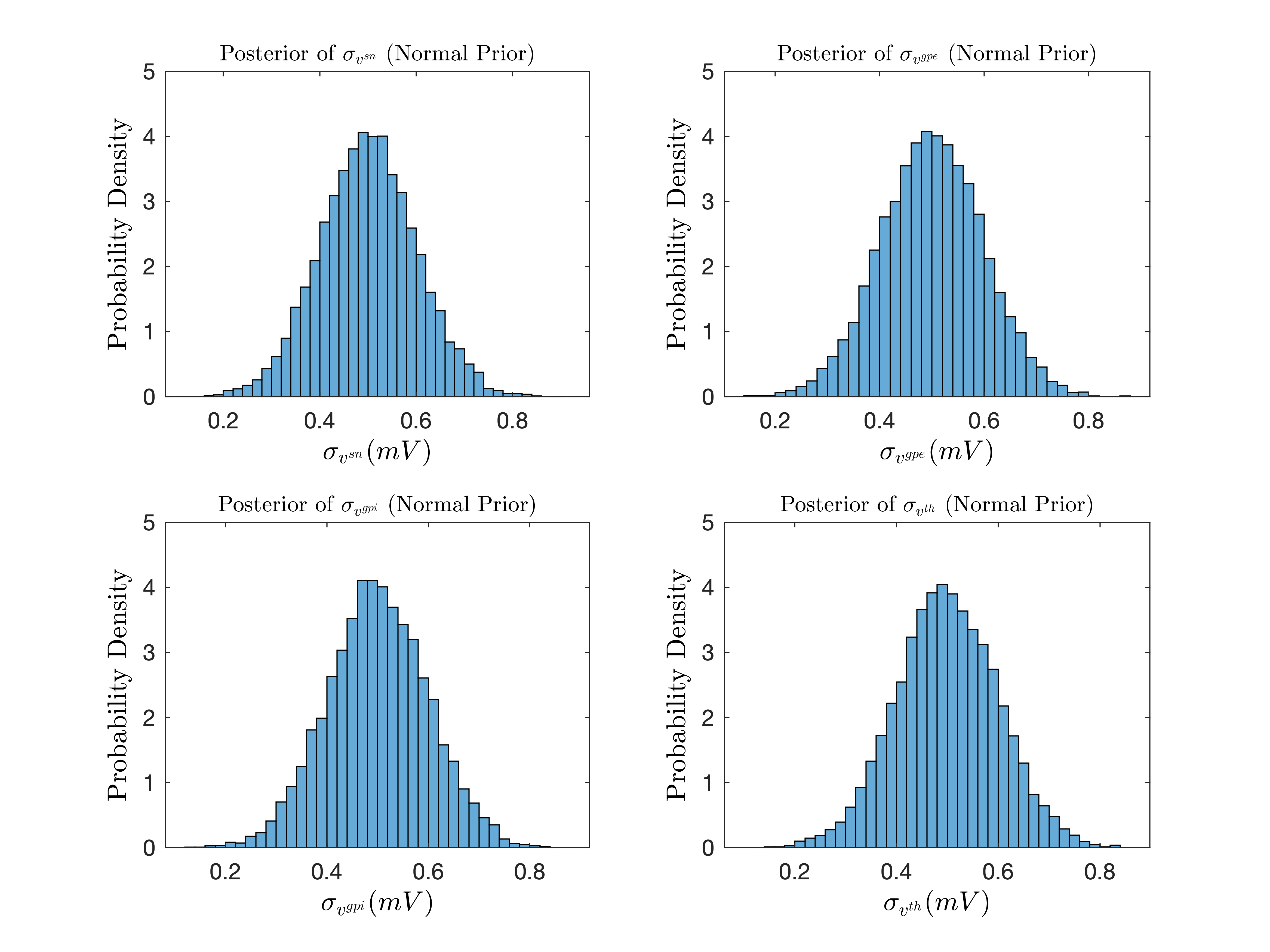}

\caption{(Color online) The posterior distributions of the standard deviations of the membrane potentials ($\sigma$) for the STN, GPe, GPi and TH, derived using Approximate Bayesian Computation (ABC).} \label{fig4A3}
\end{figure}
Finally, in Fig.~\ref{fig4A3}, each histogram represents the posterior probability density for the standard deviation of the membrane potential ($\sigma$) in each region. These distributions are centred around a mean value between $0.4 mV$ and $0.6 mV$, with relatively symmetric shapes, indicating that the uncertainty in the estimated standard deviations is well-captured and follows a normal-like distribution based on the chosen prior. The tails of these distributions taper off smoothly, which indicates that extreme values for the standard deviation are unlikely, providing confidence in the reliability of the estimated noise levels. In general, ABC allows us to estimate these parameters even when the likelihood is intractable, making it a powerful tool in this context where traditional statistical methods may fall short due to the complexity of the system. Moreover, the posterior distributions give insight into the variability of the membrane potential in the presence of noise and DBS, and by using a Bayesian framework, we are not only estimating the mean value of $\sigma$ but also quantifying the uncertainty around these estimates. This makes the results more interpretable in terms of probabilistic inference, which is a significant advantage over point estimates provided by classical methods. The standard deviations $\sigma$ estimated in Fig.~\ref{fig4A3}  reflect the degree of variability in membrane potential due to noise in each of the key regions of the basal ganglia. The fact that the posterior distributions are centred around moderate values suggests that the noise levels across these regions are relatively consistent, further supporting the idea that noise is an inherent part of the system’s dynamics. This insight can be crucial for understanding how DBS modulates the activity of these regions by reducing pathological variability while still allowing for normal physiological fluctuations. This application of ABC is a novel contribution to the study of brain network dynamics, particularly in the context of the basal ganglia and NDDs like PD. By using this approach, you provide a probabilistic framework to better understand the role of noise in the membrane potential dynamics under DBS, offering a more refined view of how DBS stabilizes the system. Moreover, the ability to infer the posterior distributions of $\sigma$ allows for the exploration of individual differences in neural responses to DBS, which can have implications for personalized treatments and tuning of DBS parameters in clinical settings.

\section{Discussion}\label{SE4}
In the current work, we used resting-state functional connectomes and machine-learning approaches such as parallel computing in classifying brain connectomes in healthy and PD states, with and without the stochastic process. To solve the network model computationally, we utilized the Euler–Maruyama method with a time-step $dt= 0.001$, with consistent results across various time-step values. Throughout all simulations, we used brain connectome data sourced from \url{https://braingraph.org}, with no significant changes observed over time.

Additionally, obtaining precise data regarding cortical-BGTH tractography proves challenging due to various limitations in structural MRI data, as highlighted by Meier et al \cite{meier2022virtual}. Petersen et al. have recently introduced an advanced axonal pathway atlas for the human brain, integrating findings from histological studies, imaging data, and expert insights \cite{abos2017discriminating,petersen2019holographic}. Earlier work optimized connection probabilities and weights among BG regions to align with empirical fMRI data on an individual basis \cite{shaheen2022multiscale}. However, many studies resort to normative connectome atlases due to the complexities associated with acquiring and interpreting patient-specific diffusion-weighted imaging data. Yet, the potential benefits of patient-specific connection data remain uncertain.

In this work, we modified the Rubin and Terman model to better align with experimental evidence on neuron firing characteristics \cite{shaheen2022multiscale,rubin2004high}. Our study focuses on a network mathematical model that enables experimentalists to quickly evaluate membrane potentials throughout the cortex's four central nuclei and BGTH regions. Importantly, significant connection changes were detected in the PD brain, especially in stochastic noise, consistent with previous findings \cite{wozniak2020deep,zheng2023noise,touboul2020noise}. Unlike prior studies focusing on group differences, our study examined the discriminatory potential of resting-state functional connectivity at the individual level. This provides evidence that connectivity patterns with stochastic noise can distinguish PD patients with cognitive impairment from those without. Notably, our study is the sole one to show this capacity. In PD patients, functional connectivity reductions were observed across significant brain regions, with a disproportionate involvement of occipital-temporal and occipital-frontal connections compared to healthy controls. These findings contribute to the understanding of PD-associated cognitive impairment, corroborating previous neuroimaging modalities' observations. 

Bayesian inference plays a pivotal role in PD research by providing a robust framework for integrating various sources of uncertainty, both in diagnosis and in modelling the disease’s progression \cite{young2024data}. In particular, Bayesian models can incorporate prior knowledge—such as the onset age, genetic factors, and early-stage symptoms—to improve diagnostic accuracy \cite{shaheen2023bayesian}. These models can be updated as new data becomes available, ensuring the predictions remain current and relevant \cite{shaheen2023data,beaumont2019approximate}. Importantly, our study highlights the effect of Bayesian inference on stochastic modelling, which reveals that the posterior distribution not only provides a more robust estimate of model parameters but also incorporates uncertainty in a principled manner \cite{ghahramani2013bayesian}. This allows for a more accurate reflection of the underlying biological processes, particularly in the presence of noise or incomplete data. The posterior distribution serves as a critical tool for refining predictions, improving the model’s fit, and guiding decision-making by manipulating prior knowledge and observed data. Ultimately, Bayesian inference enhances the interpretability and reliability of stochastic models, offering deeper insights into the dynamics of complex systems such as brain networks in NDDs \cite{dadu2023ml}.

Furthermore, this study underscores the potential of resting-state functional connectivity measures for individual-level discrimination in PD, providing valuable insights into the disease's pathophysiology. Our examination of stochastic noise's influence on various brain regions, including the STN, GPi, GPe, and thalamus, enhances our understanding of the complexities of PD. Moreover, noise can modulate the excitability of neurons within the brain network, influencing their firing patterns and synchronization properties. This modulation could change the overall network dynamics, affecting the balance between inhibitory and excitatory signals and potentially leading to dysregulated activity associated with neurological disorders such as PD. This discovery not only calls into question existing therapy options but also demonstrates the potential of our hybrid modelling approach for identifying subtle elements of brain dysfunction. Overall, investigating the influence of noise on the CBTH system using a discrete brain network paradigm gives valuable information on the system's resilience, flexibility, and susceptibility to dysfunction. It provides insights into the processes underpinning neurological diseases and may aid in developing therapeutic approaches to restore normal network function.
\section{Conclusions}\label{SE5}
In conclusion, this study presents a novel approach utilizing a fusion of ML, stochastic modelling, and connectomic data to delve into the intricate neural pathways implicated in Parkinson's disease (PD) pathogenesis. By harnessing modern computational methodologies, we've endeavoured to decode the nuanced changes in structure and function within the PD-afflicted brain. Our findings shed light on the subtle alterations in neuronal activity patterns associated with PD progression, illuminating potential targets for therapeutic intervention. The hybrid modelling framework and innovative co-simulation technique developed in this research offer a deeper understanding of the impact of stochastic disturbances on the CBGTH network within the context of large-scale brain connectivity maps derived from the HCP. Notably, our analysis reveals that even in the presence of DBS, stochastic influences can lead to heightened activity in the thalamus, a key node in PD pathology. Also, the effect of stochastic noises in brain regions, such as STN, GPe, GPi and TH, has been analyzed. This study's novelty lies in applying inference-based integrative Bayesian approaches to the developed model. Our analysis revealed that the posterior distributions of the standard deviations of the membrane potentials follow a normal distribution. Bayesian inference further provides a probabilistic framework that enhances understanding of the role of noise in membrane potential dynamics under DBS. This approach offers a more nuanced perspective on how DBS stabilizes the system by accounting for the variability in membrane potentials.

In the future, we aim to analyze high temporal and spatially resolved cerebral data sources from functional
near-infrared spectroscopy and EEG, PET, and MRI/fMRI data from healthy patients
with neurodegenerative conditions such as PD. This work lays the groundwork for novel therapeutic strategies tailored to individual patients by elucidating the complex dynamics of neuronal activity underlying PD. The integration of ML, stochastic modelling, and connectomic data holds promise for advancing our understanding of PD pathophysiology and accelerating the development of personalized treatment approaches. Ultimately, the goal is to translate these insights into tangible clinical benefits, offering hope to those affected by this debilitating neurological disorder.

\section*{Acknowledgements}
The authors are grateful to the NSERC and the CRC Program for their support. This research was enabled in part by support provided by SHARCNET \url{(www. sharcnet.ca)} and Digital Research Alliance of Canada \url{(www.alliancecan.ca)}.

\section*{Declarations}

\section*{Ethical Approval}
The authors declare that the research was conducted in the absence of any commercial or financial relationships that could be construed as a potential conflict of interest.
\section*{Competing interests}
The authors declare no competing interests.
\section*{Authors' contributions}
HS: methods and materials, data curation, formal analysis, investigation, writing, and original draft preparation. RM: conceptualization, supervision, and reviews. All authors approved the final submitted version.
\section*{Funding}
The authors are grateful to the NSERC and the CRC Program for their support. This research was enabled in part by support provided by SHARCNET \url{(www. sharcnet.ca)} and Digital Research Alliance of Canada \url{(www.alliancecan.ca)}.


\begin{thebibliography}{41}
\expandafter\ifx\csname natexlab\endcsname\relax\def\natexlab#1{#1}\fi
\providecommand{\url}[1]{\texttt{#1}}
\providecommand{\href}[2]{#2}
\providecommand{\path}[1]{#1}
\providecommand{\DOIprefix}{doi:}
\providecommand{\ArXivprefix}{arXiv:}
\providecommand{\URLprefix}{URL: }
\providecommand{\Pubmedprefix}{pmid:}
\providecommand{\doi}[1]{\href{http://dx.doi.org/#1}{\path{#1}}}
\providecommand{\Pubmed}[1]{\href{pmid:#1}{\path{#1}}}
\providecommand{\bibinfo}[2]{#2}
\ifx\xfnm\relax \def\xfnm[#1]{\unskip,\space#1}\fi
\bibitem[{Ab{\'o}s et~al.(2017)Ab{\'o}s, Baggio, Segura, Garc{\'\i}a-D{\'\i}az,
  Compta, Mart{\'\i}, Valldeoriola and Junqu{\'e}}]{abos2017discriminating}
\bibinfo{author}{Ab{\'o}s, A.}, \bibinfo{author}{Baggio, H.C.},
  \bibinfo{author}{Segura, B.}, \bibinfo{author}{Garc{\'\i}a-D{\'\i}az, A.I.},
  \bibinfo{author}{Compta, Y.}, \bibinfo{author}{Mart{\'\i}, M.J.},
  \bibinfo{author}{Valldeoriola, F.}, \bibinfo{author}{Junqu{\'e}, C.},
  \bibinfo{year}{2017}.
\newblock \bibinfo{title}{Discriminating cognitive status in parkinson’s
  disease through functional connectomics and machine learning}.
\newblock \bibinfo{journal}{Scientific reports} \bibinfo{volume}{7},
  \bibinfo{pages}{45347}.
\bibitem[{Beaumont(2019)}]{beaumont2019approximate}
\bibinfo{author}{Beaumont, M.A.}, \bibinfo{year}{2019}.
\newblock \bibinfo{title}{Approximate bayesian computation}.
\newblock \bibinfo{journal}{Annual Review of Statistics and its Application}
  \bibinfo{volume}{6}, \bibinfo{pages}{379--403}.
\bibitem[{Carron et~al.(2013)Carron, Chaillet, Filipchuk, Pasillas-L{\'e}pine
  and Hammond}]{carron2013closing}
\bibinfo{author}{Carron, R.}, \bibinfo{author}{Chaillet, A.},
  \bibinfo{author}{Filipchuk, A.}, \bibinfo{author}{Pasillas-L{\'e}pine, W.},
  \bibinfo{author}{Hammond, C.}, \bibinfo{year}{2013}.
\newblock \bibinfo{title}{Closing the loop of deep brain stimulation}.
\newblock \bibinfo{journal}{Frontiers in systems neuroscience}
  \bibinfo{volume}{7}, \bibinfo{pages}{112}.
\bibitem[{Charalambous and Djebbara(2023)}]{charalambous2023natural}
\bibinfo{author}{Charalambous, E.}, \bibinfo{author}{Djebbara, Z.},
  \bibinfo{year}{2023}.
\newblock \bibinfo{title}{On natural attunement: shared rhythms between the
  brain and the environment}.
\newblock \bibinfo{journal}{Neuroscience \& Biobehavioral Reviews}
  \bibinfo{volume}{155}, \bibinfo{pages}{105438}.
\bibitem[{Dadu(2023)}]{dadu2023ml}
\bibinfo{author}{Dadu, A.}, \bibinfo{year}{2023}.
\newblock \bibinfo{title}{ML-assisted therapeutics for neurodegenerative
  disorders}.
\newblock Ph.D. thesis. University of Illinois at Urbana-Champaign.
\bibitem[{Deco et~al.(2009)Deco, Rolls and Romo}]{deco2009stochastic}
\bibinfo{author}{Deco, G.}, \bibinfo{author}{Rolls, E.T.},
  \bibinfo{author}{Romo, R.}, \bibinfo{year}{2009}.
\newblock \bibinfo{title}{Stochastic dynamics as a principle of brain
  function}.
\newblock \bibinfo{journal}{Progress in neurobiology} \bibinfo{volume}{88},
  \bibinfo{pages}{1--16}.
\bibitem[{Deffains et~al.(2016)Deffains, Iskhakova, Katabi, Haber, Israel and
  Bergman}]{deffains2016subthalamic}
\bibinfo{author}{Deffains, M.}, \bibinfo{author}{Iskhakova, L.},
  \bibinfo{author}{Katabi, S.}, \bibinfo{author}{Haber, S.N.},
  \bibinfo{author}{Israel, Z.}, \bibinfo{author}{Bergman, H.},
  \bibinfo{year}{2016}.
\newblock \bibinfo{title}{Subthalamic, not striatal, activity correlates with
  basal ganglia downstream activity in normal and parkinsonian monkeys}.
\newblock \bibinfo{journal}{Elife} \bibinfo{volume}{5},
  \bibinfo{pages}{e16443}.
\bibitem[{Ghahramani(2013)}]{ghahramani2013bayesian}
\bibinfo{author}{Ghahramani, Z.}, \bibinfo{year}{2013}.
\newblock \bibinfo{title}{Bayesian non-parametrics and the probabilistic
  approach to modelling}.
\newblock \bibinfo{journal}{Philosophical Transactions of the Royal Society A:
  Mathematical, Physical and Engineering Sciences} \bibinfo{volume}{371},
  \bibinfo{pages}{20110553}.
\bibitem[{Ghebrehiwet et~al.(2024)Ghebrehiwet, Zaki, Damseh and
  Mohamad}]{ghebrehiwet2024revolutionizing}
\bibinfo{author}{Ghebrehiwet, I.}, \bibinfo{author}{Zaki, N.},
  \bibinfo{author}{Damseh, R.}, \bibinfo{author}{Mohamad, M.S.},
  \bibinfo{year}{2024}.
\newblock \bibinfo{title}{Revolutionizing personalized medicine with generative
  ai: a systematic review}.
\newblock \bibinfo{journal}{Artificial Intelligence Review}
  \bibinfo{volume}{57}, \bibinfo{pages}{1--41}.
\bibitem[{Gutkin et~al.(2005)Gutkin, Ermentrout and Reyes}]{gutkin2005phase}
\bibinfo{author}{Gutkin, B.S.}, \bibinfo{author}{Ermentrout, G.B.},
  \bibinfo{author}{Reyes, A.D.}, \bibinfo{year}{2005}.
\newblock \bibinfo{title}{Phase-response curves give the responses of neurons
  to transient inputs}.
\newblock \bibinfo{journal}{Journal of neurophysiology} \bibinfo{volume}{94},
  \bibinfo{pages}{1623--1635}.
\bibitem[{Johnson et~al.(2024)Johnson, Okun, Scangos, Mayberg and
  de~Hemptinne}]{johnson2024deep}
\bibinfo{author}{Johnson, K.A.}, \bibinfo{author}{Okun, M.S.},
  \bibinfo{author}{Scangos, K.W.}, \bibinfo{author}{Mayberg, H.S.},
  \bibinfo{author}{de~Hemptinne, C.}, \bibinfo{year}{2024}.
\newblock \bibinfo{title}{Deep brain stimulation for refractory major
  depressive disorder: a comprehensive review}.
\newblock \bibinfo{journal}{Molecular Psychiatry} , \bibinfo{pages}{1--13}.
\bibitem[{Kim et~al.(2017)Kim, Criaud, Cho, D{\'\i}ez-Cirarda, Mihaescu,
  Coakeley, Ghadery, Valli, Jacobs, Houle et~al.}]{kim2017abnormal}
\bibinfo{author}{Kim, J.}, \bibinfo{author}{Criaud, M.}, \bibinfo{author}{Cho,
  S.S.}, \bibinfo{author}{D{\'\i}ez-Cirarda, M.}, \bibinfo{author}{Mihaescu,
  A.}, \bibinfo{author}{Coakeley, S.}, \bibinfo{author}{Ghadery, C.},
  \bibinfo{author}{Valli, M.}, \bibinfo{author}{Jacobs, M.F.},
  \bibinfo{author}{Houle, S.}, et~al., \bibinfo{year}{2017}.
\newblock \bibinfo{title}{Abnormal intrinsic brain functional network dynamics
  in parkinson’s disease}.
\newblock \bibinfo{journal}{Brain} \bibinfo{volume}{140},
  \bibinfo{pages}{2955--2967}.
\bibitem[{Liang et~al.(2024)Liang, Yang and Zhou}]{liang2024excitation}
\bibinfo{author}{Liang, J.}, \bibinfo{author}{Yang, Z.}, \bibinfo{author}{Zhou,
  C.}, \bibinfo{year}{2024}.
\newblock \bibinfo{title}{Excitation--inhibition balance, neural criticality,
  and activities in neuronal circuits}.
\newblock \bibinfo{journal}{The Neuroscientist} ,
  \bibinfo{pages}{10738584231221766}.
\bibitem[{Liu et~al.(2018)Liu, Wang, Deng, Li, Fietkiewicz and
  Loparo}]{liu2018noise}
\bibinfo{author}{Liu, C.}, \bibinfo{author}{Wang, J.}, \bibinfo{author}{Deng,
  B.}, \bibinfo{author}{Li, H.}, \bibinfo{author}{Fietkiewicz, C.},
  \bibinfo{author}{Loparo, K.A.}, \bibinfo{year}{2018}.
\newblock \bibinfo{title}{Noise-induced improvement of the parkinsonian state:
  A computational study}.
\newblock \bibinfo{journal}{IEEE Transactions on Cybernetics}
  \bibinfo{volume}{49}, \bibinfo{pages}{3655--3664}.
\bibitem[{Meier et~al.(2022)Meier, Perdikis, Blickensd{\"o}rfer, Stefanovski,
  Liu, Maith, Dinkelbach, Baladron, Hamker and Ritter}]{meier2022virtual}
\bibinfo{author}{Meier, J.M.}, \bibinfo{author}{Perdikis, D.},
  \bibinfo{author}{Blickensd{\"o}rfer, A.}, \bibinfo{author}{Stefanovski, L.},
  \bibinfo{author}{Liu, Q.}, \bibinfo{author}{Maith, O.},
  \bibinfo{author}{Dinkelbach, H.{\"U}.}, \bibinfo{author}{Baladron, J.},
  \bibinfo{author}{Hamker, F.H.}, \bibinfo{author}{Ritter, P.},
  \bibinfo{year}{2022}.
\newblock \bibinfo{title}{Virtual deep brain stimulation: Multiscale
  co-simulation of a spiking basal ganglia model and a whole-brain mean-field
  model with the virtual brain}.
\newblock \bibinfo{journal}{Experimental Neurology} \bibinfo{volume}{354},
  \bibinfo{pages}{114111}.
\bibitem[{Novelli et~al.(2024)Novelli, Friston and Razi}]{novelli2024spectral}
\bibinfo{author}{Novelli, L.}, \bibinfo{author}{Friston, K.},
  \bibinfo{author}{Razi, A.}, \bibinfo{year}{2024}.
\newblock \bibinfo{title}{Spectral dynamic causal modeling: A didactic
  introduction and its relationship with functional connectivity}.
\newblock \bibinfo{journal}{Network Neuroscience} \bibinfo{volume}{8},
  \bibinfo{pages}{178--202}.
\bibitem[{Oliveira et~al.(2023)Oliveira, Coelho, Carvalho, Ferreira-Pinto, Vaz
  and Aguiar}]{oliveira2023machine}
\bibinfo{author}{Oliveira, A.M.}, \bibinfo{author}{Coelho, L.},
  \bibinfo{author}{Carvalho, E.}, \bibinfo{author}{Ferreira-Pinto, M.J.},
  \bibinfo{author}{Vaz, R.}, \bibinfo{author}{Aguiar, P.},
  \bibinfo{year}{2023}.
\newblock \bibinfo{title}{Machine learning for adaptive deep brain stimulation
  in parkinson’s disease: closing the loop}.
\newblock \bibinfo{journal}{Journal of Neurology} \bibinfo{volume}{270},
  \bibinfo{pages}{5313--5326}.
\bibitem[{Peralta et~al.(2021)Peralta, Jannin and Baxter}]{peralta2021machine}
\bibinfo{author}{Peralta, M.}, \bibinfo{author}{Jannin, P.},
  \bibinfo{author}{Baxter, J.S.}, \bibinfo{year}{2021}.
\newblock \bibinfo{title}{Machine learning in deep brain stimulation: A
  systematic review}.
\newblock \bibinfo{journal}{Artificial Intelligence in Medicine}
  \bibinfo{volume}{122}, \bibinfo{pages}{102198}.
\bibitem[{Petersen et~al.(2019)Petersen, Mlakar, Haber, Parent, Smith, Strick,
  Griswold and McIntyre}]{petersen2019holographic}
\bibinfo{author}{Petersen, M.V.}, \bibinfo{author}{Mlakar, J.},
  \bibinfo{author}{Haber, S.N.}, \bibinfo{author}{Parent, M.},
  \bibinfo{author}{Smith, Y.}, \bibinfo{author}{Strick, P.L.},
  \bibinfo{author}{Griswold, M.A.}, \bibinfo{author}{McIntyre, C.C.},
  \bibinfo{year}{2019}.
\newblock \bibinfo{title}{Holographic reconstruction of axonal pathways in the
  human brain}.
\newblock \bibinfo{journal}{Neuron} \bibinfo{volume}{104},
  \bibinfo{pages}{1056--1064}.
\bibitem[{Rubin and Terman(2004)}]{rubin2004high}
\bibinfo{author}{Rubin, J.E.}, \bibinfo{author}{Terman, D.},
  \bibinfo{year}{2004}.
\newblock \bibinfo{title}{High frequency stimulation of the subthalamic nucleus
  eliminates pathological thalamic rhythmicity in a computational model}.
\newblock \bibinfo{journal}{Journal of computational neuroscience}
  \bibinfo{volume}{16}, \bibinfo{pages}{211--235}.
\bibitem[{Salaramoli et~al.(2024)Salaramoli, Joshaghani, Hosseini and
  Hashemy}]{salaramoli2024therapeutic}
\bibinfo{author}{Salaramoli, S.}, \bibinfo{author}{Joshaghani, H.R.},
  \bibinfo{author}{Hosseini, M.}, \bibinfo{author}{Hashemy, S.I.},
  \bibinfo{year}{2024}.
\newblock \bibinfo{title}{Therapeutic effects of selenium on alpha-synuclein
  accumulation in substantia nigra pars compacta in a rat model of
  parkinson’s disease: Behavioral and biochemical outcomes}.
\newblock \bibinfo{journal}{Biological trace element research}
  \bibinfo{volume}{202}, \bibinfo{pages}{1115--1125}.
\bibitem[{Seguin et~al.(2023a)Seguin, Jedynak, David, Mansour, Sporns and
  Zalesky}]{seguin2023communication}
\bibinfo{author}{Seguin, C.}, \bibinfo{author}{Jedynak, M.},
  \bibinfo{author}{David, O.}, \bibinfo{author}{Mansour, S.},
  \bibinfo{author}{Sporns, O.}, \bibinfo{author}{Zalesky, A.},
  \bibinfo{year}{2023}a.
\newblock \bibinfo{title}{Communication dynamics in the human connectome shape
  the cortex-wide propagation of direct electrical stimulation}.
\newblock \bibinfo{journal}{Neuron} \bibinfo{volume}{111},
  \bibinfo{pages}{1391--1401}.
\bibitem[{Seguin et~al.(2023b)Seguin, Sporns and Zalesky}]{seguin2023brain}
\bibinfo{author}{Seguin, C.}, \bibinfo{author}{Sporns, O.},
  \bibinfo{author}{Zalesky, A.}, \bibinfo{year}{2023}b.
\newblock \bibinfo{title}{Brain network communication: concepts, models and
  applications}.
\newblock \bibinfo{journal}{Nature reviews neuroscience} \bibinfo{volume}{24},
  \bibinfo{pages}{557--574}.
\bibitem[{Shaheen and Melnik(2022)}]{shaheen2022deep}
\bibinfo{author}{Shaheen, H.}, \bibinfo{author}{Melnik, R.},
  \bibinfo{year}{2022}.
\newblock \bibinfo{title}{Deep brain stimulation with a computational model for
  the cortex-thalamus-basal-ganglia system and network dynamics of neurological
  disorders}.
\newblock \bibinfo{journal}{Computational and Mathematical Methods}
  \bibinfo{volume}{2022}, \bibinfo{pages}{8998150}.
\bibitem[{Shaheen and Melnik(2024)}]{shaheen2024neural}
\bibinfo{author}{Shaheen, H.}, \bibinfo{author}{Melnik, R.},
  \bibinfo{year}{2024}.
\newblock \bibinfo{title}{Neural dynamics in parkinson’s disease: Integrating
  machine learning and stochastic modelling with connectomic data}, in:
  \bibinfo{booktitle}{International Conference on Computational Science},
  \bibinfo{organization}{Springer}. pp. \bibinfo{pages}{46--60}.
\bibitem[{Shaheen et~al.(2023)Shaheen, Melnik, Initiative
  et~al.}]{shaheen2023bayesian}
\bibinfo{author}{Shaheen, H.}, \bibinfo{author}{Melnik, R.},
  \bibinfo{author}{Initiative, A.D.N.}, et~al., \bibinfo{year}{2023}.
\newblock \bibinfo{title}{Bayesian inference and role of astrocytes in
  amyloid-beta dynamics with modelling of alzheimer's disease using clinical
  data}.
\newblock \bibinfo{journal}{arXiv e-prints} , \bibinfo{pages}{arXiv--2306}.
\bibitem[{Shaheen et~al.(2024)Shaheen, Melnik and Sundeep}]{shaheen2023data}
\bibinfo{author}{Shaheen, H.}, \bibinfo{author}{Melnik, R.},
  \bibinfo{author}{Sundeep, S.}, \bibinfo{year}{2024}.
\newblock \bibinfo{title}{Data-driven stochastic model for quantifying the
  interplay between amyloid-beta and calcium levels in alzheimer's disease}.
\newblock \bibinfo{journal}{Statistical Analysis and Data Mining: The ASA Data
  Science Journal} \bibinfo{volume}{17}, \bibinfo{pages}{e11679}.
\bibitem[{Shaheen et~al.(2022)Shaheen, Pal and Melnik}]{shaheen2022multiscale}
\bibinfo{author}{Shaheen, H.}, \bibinfo{author}{Pal, S.},
  \bibinfo{author}{Melnik, R.}, \bibinfo{year}{2022}.
\newblock \bibinfo{title}{Multiscale co-simulation of deep brain stimulation
  with brain networks in neurodegenerative disorders}.
\newblock \bibinfo{journal}{Brain Multiphysics} \bibinfo{volume}{3},
  \bibinfo{pages}{100058}.
\bibitem[{Shi et~al.(2023)Shi, Li, Zhang, Li and Han}]{shi2023characteristic}
\bibinfo{author}{Shi, P.}, \bibinfo{author}{Li, J.}, \bibinfo{author}{Zhang,
  W.}, \bibinfo{author}{Li, M.}, \bibinfo{author}{Han, D.},
  \bibinfo{year}{2023}.
\newblock \bibinfo{title}{Characteristic frequency detection of steady-state
  visual evoked potentials based on filter bank second-order underdamped
  tristable stochastic resonance}.
\newblock \bibinfo{journal}{Biomedical Signal Processing and Control}
  \bibinfo{volume}{84}, \bibinfo{pages}{104817}.
\bibitem[{Staffaroni et~al.(2019)Staffaroni, Cobigo, Elahi, Casaletto, Walters,
  Wolf, Lindbergh, Rosen and Kramer}]{staffaroni2019longitudinal}
\bibinfo{author}{Staffaroni, A.M.}, \bibinfo{author}{Cobigo, Y.},
  \bibinfo{author}{Elahi, F.M.}, \bibinfo{author}{Casaletto, K.B.},
  \bibinfo{author}{Walters, S.M.}, \bibinfo{author}{Wolf, A.},
  \bibinfo{author}{Lindbergh, C.A.}, \bibinfo{author}{Rosen, H.J.},
  \bibinfo{author}{Kramer, J.H.}, \bibinfo{year}{2019}.
\newblock \bibinfo{title}{A longitudinal characterization of perfusion in the
  aging brain and associations with cognition and neural structure}.
\newblock \bibinfo{journal}{Human brain mapping} \bibinfo{volume}{40},
  \bibinfo{pages}{3522--3533}.
\bibitem[{Storm et~al.(2024)Storm, Klink, Aru, Senn, Goebel, Pigorini,
  Avanzini, Vanduffel, Roelfsema, Massimini et~al.}]{storm2024integrative}
\bibinfo{author}{Storm, J.F.}, \bibinfo{author}{Klink, P.C.},
  \bibinfo{author}{Aru, J.}, \bibinfo{author}{Senn, W.},
  \bibinfo{author}{Goebel, R.}, \bibinfo{author}{Pigorini, A.},
  \bibinfo{author}{Avanzini, P.}, \bibinfo{author}{Vanduffel, W.},
  \bibinfo{author}{Roelfsema, P.R.}, \bibinfo{author}{Massimini, M.}, et~al.,
  \bibinfo{year}{2024}.
\newblock \bibinfo{title}{An integrative, multiscale view on neural theories of
  consciousness}.
\newblock \bibinfo{journal}{Neuron} \bibinfo{volume}{112},
  \bibinfo{pages}{1531--1552}.
\bibitem[{Sumarac et~al.(2024)Sumarac, Youn, Fearon, Zivkovic, Keerthi, Flouty,
  Popovic, Hodaie, Kalia, Lozano et~al.}]{sumarac2024clinico}
\bibinfo{author}{Sumarac, S.}, \bibinfo{author}{Youn, J.},
  \bibinfo{author}{Fearon, C.}, \bibinfo{author}{Zivkovic, L.},
  \bibinfo{author}{Keerthi, P.}, \bibinfo{author}{Flouty, O.},
  \bibinfo{author}{Popovic, M.}, \bibinfo{author}{Hodaie, M.},
  \bibinfo{author}{Kalia, S.}, \bibinfo{author}{Lozano, A.}, et~al.,
  \bibinfo{year}{2024}.
\newblock \bibinfo{title}{Clinico-physiological correlates of parkinson’s
  disease from multi-resolution basal ganglia recordings}.
\newblock \bibinfo{journal}{npj Parkinson's Disease} \bibinfo{volume}{10},
  \bibinfo{pages}{175}.
\bibitem[{Tai et~al.(2019)Tai, Albuquerque, Carmona, Subramanieapillai, Cha,
  Sheko, Lee, Mansur and McIntyre}]{tai2019machine}
\bibinfo{author}{Tai, A.M.}, \bibinfo{author}{Albuquerque, A.},
  \bibinfo{author}{Carmona, N.E.}, \bibinfo{author}{Subramanieapillai, M.},
  \bibinfo{author}{Cha, D.S.}, \bibinfo{author}{Sheko, M.},
  \bibinfo{author}{Lee, Y.}, \bibinfo{author}{Mansur, R.},
  \bibinfo{author}{McIntyre, R.S.}, \bibinfo{year}{2019}.
\newblock \bibinfo{title}{Machine learning and big data: Implications for
  disease modeling and therapeutic discovery in psychiatry}.
\newblock \bibinfo{journal}{Artificial intelligence in medicine}
  \bibinfo{volume}{99}, \bibinfo{pages}{101704}.
\bibitem[{Thieu and Melnik(2022)}]{thieu2022coupled}
\bibinfo{author}{Thieu, T.K.T.}, \bibinfo{author}{Melnik, R.},
  \bibinfo{year}{2022}.
\newblock \bibinfo{title}{Coupled effects of channels and synaptic dynamics in
  stochastic modelling of healthy and parkinson's-disease-affected brains.}
\newblock \bibinfo{journal}{AIMS Bioengineering} \bibinfo{volume}{9}.
\bibitem[{Touboul et~al.(2020)Touboul, Piette, Venance and
  Ermentrout}]{touboul2020noise}
\bibinfo{author}{Touboul, J.D.}, \bibinfo{author}{Piette, C.},
  \bibinfo{author}{Venance, L.}, \bibinfo{author}{Ermentrout, G.B.},
  \bibinfo{year}{2020}.
\newblock \bibinfo{title}{Noise-induced synchronization and antiresonance in
  interacting excitable systems: applications to deep brain stimulation in
  parkinson’s disease}.
\newblock \bibinfo{journal}{Physical Review X} \bibinfo{volume}{10},
  \bibinfo{pages}{011073}.
\bibitem[{Tyralis and Papacharalampous(2024)}]{tyralis2024review}
\bibinfo{author}{Tyralis, H.}, \bibinfo{author}{Papacharalampous, G.},
  \bibinfo{year}{2024}.
\newblock \bibinfo{title}{A review of predictive uncertainty estimation with
  machine learning}.
\newblock \bibinfo{journal}{Artificial Intelligence Review}
  \bibinfo{volume}{57}, \bibinfo{pages}{94}.
\bibitem[{Vashistha et~al.(2024)Vashistha, Moradi, Hammond, O’Brien,
  Rominger, Sari, Shi, Vegh and Reutens}]{vashistha2024parapet}
\bibinfo{author}{Vashistha, R.}, \bibinfo{author}{Moradi, H.},
  \bibinfo{author}{Hammond, A.}, \bibinfo{author}{O’Brien, K.},
  \bibinfo{author}{Rominger, A.}, \bibinfo{author}{Sari, H.},
  \bibinfo{author}{Shi, K.}, \bibinfo{author}{Vegh, V.},
  \bibinfo{author}{Reutens, D.}, \bibinfo{year}{2024}.
\newblock \bibinfo{title}{Parapet: non-invasive deep learning method for direct
  parametric brain pet reconstruction using histoimages}.
\newblock \bibinfo{journal}{EJNMMI research} \bibinfo{volume}{14},
  \bibinfo{pages}{10}.
\bibitem[{Volgushev et~al.(1998)Volgushev, Chistiakova and
  Singer}]{volgushev1998modification}
\bibinfo{author}{Volgushev, M.}, \bibinfo{author}{Chistiakova, M.},
  \bibinfo{author}{Singer, W.}, \bibinfo{year}{1998}.
\newblock \bibinfo{title}{Modification of discharge patterns of neocortical
  neurons by induced oscillations of the membrane potential}.
\newblock \bibinfo{journal}{Neuroscience} \bibinfo{volume}{83},
  \bibinfo{pages}{15--25}.
\bibitem[{Wo{\'z}niak et~al.(2020)Wo{\'z}niak, Pantazi, Bohnstingl and
  Eleftheriou}]{wozniak2020deep}
\bibinfo{author}{Wo{\'z}niak, S.}, \bibinfo{author}{Pantazi, A.},
  \bibinfo{author}{Bohnstingl, T.}, \bibinfo{author}{Eleftheriou, E.},
  \bibinfo{year}{2020}.
\newblock \bibinfo{title}{Deep learning incorporating biologically inspired
  neural dynamics and in-memory computing}.
\newblock \bibinfo{journal}{Nature Machine Intelligence} \bibinfo{volume}{2},
  \bibinfo{pages}{325--336}.
\bibitem[{Young et~al.(2024)Young, Oxtoby, Garbarino, Fox, Barkhof, Schott and
  Alexander}]{young2024data}
\bibinfo{author}{Young, A.L.}, \bibinfo{author}{Oxtoby, N.P.},
  \bibinfo{author}{Garbarino, S.}, \bibinfo{author}{Fox, N.C.},
  \bibinfo{author}{Barkhof, F.}, \bibinfo{author}{Schott, J.M.},
  \bibinfo{author}{Alexander, D.C.}, \bibinfo{year}{2024}.
\newblock \bibinfo{title}{Data-driven modelling of neurodegenerative disease
  progression: thinking outside the black box}.
\newblock \bibinfo{journal}{Nature Reviews Neuroscience} \bibinfo{volume}{25},
  \bibinfo{pages}{111--130}.
\bibitem[{Zheng et~al.(2023)Zheng, Tang, Zheng, Wang, Liu, Yang, Zhen and
  Zheng}]{zheng2023noise}
\bibinfo{author}{Zheng, Y.}, \bibinfo{author}{Tang, S.},
  \bibinfo{author}{Zheng, H.}, \bibinfo{author}{Wang, X.},
  \bibinfo{author}{Liu, L.}, \bibinfo{author}{Yang, Y.}, \bibinfo{author}{Zhen,
  Y.}, \bibinfo{author}{Zheng, Z.}, \bibinfo{year}{2023}.
\newblock \bibinfo{title}{Noise improves the association between effects of
  local stimulation and structural degree of brain networks}.
\newblock \bibinfo{journal}{PLOS Computational Biology} \bibinfo{volume}{19},
  \bibinfo{pages}{e1010866}.

\end{thebibliography}
\end{document}